\date{27December, 2007}

\documentclass[12pt]{amsart}
\usepackage{latexsym,amsmath,amsfonts,amscd,amssymb}
\usepackage{graphics}
\newtheorem{theorem}{Theorem}[section]
\newtheorem{lemma}[theorem]{Lemma}
\newtheorem{proposition}[theorem]{Proposition}
\newtheorem{corollary}[theorem]{Corollary}

\newtheorem{definition}[theorem]{Definition}

\newtheorem{example}[theorem]{Example}
\newtheorem{postulate}[theorem]{Postulate}
\newtheorem{principleA}[theorem]{Principle of virtual works}
\newtheorem{nada}[theorem]{}

\theoremstyle{remark}

\theoremstyle{remarks}
\newtheorem{remarks}[theorem]{Remarks}
\theoremstyle{remarkN}
\newtheorem{remarkN}[theorem]{Remark}

\newcommand{\la}{\langle}
\newcommand{\ra}{\rangle}

\newcommand{\cC}{{\mathcal C}}
\newcommand{\cD}{{\mathcal D}}

\newcommand{\cM}{{\mathcal M}}

\newcommand{\cL}{{\mathcal L}}

\newcommand{\cR}{{\mathcal R}}

\newcommand{\cU}{{\mathcal U}}
\newcommand{\cV}{{\mathcal V}}

\renewcommand{\a}{\alpha}
\renewcommand{\b}{\beta}
\renewcommand{\d}{\delta}

\renewcommand{\l}{\lambda}

\newcommand{\s}{\sigma}
\newcommand{\m}{\mu}
\newcommand{\n}{\nu}
\renewcommand{\o}{\omega}

\renewcommand{\t}{\tau}

\renewcommand{\O}{\Omega}

\renewcommand{\L}{\Lambda}

\begin{document}

\title[The structure of time and inertial forces in Lagrangian mechanics]{The structure of time and inertial forces in Lagrangian mechanics}


\author[J. Mu\~{n}oz D\'\i az]{J. Mu\~{n}oz D\'\i az}

\address{ Universidad de Salamanca, Departamento de Matem\'aticas, Plaza de la Merced 1, 37008-Salamanca (Spain)}

\email{clint@usal.es}

\maketitle

\bigskip

{ \centerline{\sc abstract}}

\bigskip

\begin{minipage}{14cm}
Classically time is kept fixed for infinitesimal variations in
problems in mechanics. Apparently, there appears to be no
mathematical justification in the literature for this standard
procedure. This can be explained canonically by unveiling the
intrinsic mathematical structure of time in Lagrangian mechanics.
Moreover, this structure also offers a general method to deal with
inertial forces.
\end{minipage}

\tableofcontents

\setcounter{section}{-1}

\section*{Introduction}

This article studies two related questions in Lagrangian mechanics:
the ma\-the\-ma\-ti\-cal nature of time and the one of reference
frames and inertial forces.

An  isolated mechanical system without constraints corresponds to a
Riemannian manifold $(M, T_2)$ (the configuration space) and a
1-form $\a$ on $TM$ (the work form). The general abstract version of
Newton law postulates the existence of an associated tangent field
$D$ on $TM$, that is, a second order differential equation, which is
the motion law for $(M,T_2,\a)$. There is no function $f$ on $M$
that could be used as parameter for all solutions of $D$.
Classically a new function $t$ is added to the system. This is ``the
time'', which is measured by some ``clock'' out of the system. Then
$M$ is replaced by $\mathbb{R}\times M$ and to the collection of
differential equations, $\dot t=1$ is added. When constraints
(holonomic or not) are introduced in the extended manifold
$\mathbb{R}\times M$, the classical procedure to derive the
equations of motion is to apply the {\it principle of virtual works}
using infinitesimal displacements in which \emph{the time remains
unchanged}. This is usually done without justification, or providing
only ``physical'' arguments; see e.g. \cite{Juvet}, pag. 3,
\cite{Nordheim}, pag. 48-49, \cite{Prange}, pag. 522,
\cite{Sommerfeld}, pag. 65, \cite{Whittaker}, pag. 215.

 The addition of an external time to the configuration space is
somewhat artificial. In an isolated system there is no ``time''
function, but there is a canonical ``class of time'' consisting of
all the 1-forms on $TM$, out of the zero section, that contracted
with second order differential equations give $1$. When these forms
are specialized to each solution of a second order differential
equation their primitive functions serve as time parameter.

 The admissible infinitesimal variations in mechanics are described as those
tangent fields $\d$ on $TM$ which are infinitesimal contact
transformations, projecting to $M$, and preserving the class of
time. When properly formulated, this last condition appears to be
equivalent to the classical commutation formula $d\circ\d=\d\circ
d$.

 Hypersurfaces  on $TM$ equipped with local time functions are
``time constraints'' considered in Section 5. When free mechanical
system with such constraint are considered, the equations of motion
are modified in a similar way as for ordinary holonomic constraints.
In a precise sense time constraints can be proved to be deformable
to ordinary holonomic constraints. The precise mathematical
justification for keeping the time fixed in infinitesimal
displacements in D'Alembert principle is given by Theorem
\ref{modificadaNewton3}.

 Our presentation of Lagrangian mechanics provides a natural
framework for the understanding of general inertial forces. The
nature of such forces seems to be an obscure point in the classical
literature.

Given a manifold isomorphism $\varphi\colon\cR\to\cM$, where $\cR$,
$\cM$ are provided with respective pseudo-Riemannian metrics $T_2$,
$\overline T_2$, for each mechanical structure $(\cM, \overline
T_2,\overline\a)$ on $M$, there correspond \emph{two} mechanical
structures $(\cR,T_2,\a)$, $(\cR,T_2,\a_1)$ canonically associated
by $\varphi$ to the one in $\cM$.  Their difference can be read in
$\cR$ as the \emph{inertial force} caused in $(\cM,\overline
T_2,\overline\a)$ by the \emph{reference frame} $\varphi$. This
gives a precise mathematical meaning to the expression ``$\varphi$
preserves the equations of motion'' that spreads across the
literature without a proper previous definition. The property of
``preserving the equations of motion'' is interesting only when time
constraints are present, as is the case for uniparametric
automorphism groups, because for free systems, $\varphi$ preserves
the equation of motion only when it is an isometry, and therefore a
simple change of coordinates.

Sections 0, 1, 2, 3 have been included to make the article
self-contained, although the material therein is classical, except
for language.

\section{Notations and definitions} \label{sec:notations}

We start with a brief overview of notations and definitions used in
the article.

Let $M$ be a smooth manifold of dimension $n$, and $TM$ be its
tangent bundle.

Each differential $1$-form $\alpha$ on $M$ can be considered as a
function on $TM$, denoted by $\dot\alpha$, which assigns to each
$v_a\in T_a M$ the value $\dot\alpha(v_a)=\langle\alpha_a,
v_a\rangle$ obtained by duality. In particular, a function
$f\in\cC^\infty(M)$ defines the function on $TM$ associated to $df$
that we denote in short by $\dot f$. This definition also applies to
differential forms $\alpha$ on $TM$ that are at each point the
pull-back of a form on $M$. In the sequel we call these forms {\it
horizontal forms}.

The map $f\mapsto \dot f$  from $\cC^\infty(M)$ to
$\cC^\infty(TM)$ is a derivation of the ring $\cC^\infty(M)$
taking values in the $\cC^\infty(M)$-module $\cC^\infty(TM)$. We
denote it by $\dot d$ since it is essentially the differential.
For each horizontal form $\a$, we have $\dot\a=\la\a, \dot d\ra$
as functions on $TM$.

Any derivation $\delta$ of $\cC^\infty(M)$ to the
$\cC^\infty(M)$-module $\cC^\infty(TM)$ can be viewed as a
\emph{field in $TM$ taking values in $TM$}, that is, as a rule
assigning to each point $v_a\in T_aM$ a tangent vector
$\delta_{v_a}\in T_aM$. More precisely, $\delta_{v_a}f=(\delta
f)(v_a)$, for each $f\in\cC^\infty(M)$. In particular, the
derivation $\dot d$ is the identity vector field in $TM$,
$v_a\mapsto v_a$, since $\dot d_{v_a}f=(\dot d f)(v_a)=\langle df,
v_a\rangle=v_a(f)$.

Using the vector space structure of the fibers of $TM$ we can
associate to each $v_a\in T_aM$ a tangent vector to $T_aM$ at each
point as the derivative along $v_a$ in $T_aM$. Denoting by $V_a$
this derivation, we have for $f\in\cC^\infty(M)$ and a point $w_a\in
T_aM$:
\begin{equation*}
 V_a(\dot f)(w_a)=\lim_{t\rightarrow 0}
              \frac{\dot f(w_a+tv_a)-\dot f(w_a)}{t}=\dot
              f(v_a)=v_a(f)\ .
 \end{equation*}

At each $w_a\in T_aM$,  $V_a\in T_{w_a}(T_aM)$ is called the
\emph{vertical representative} of  $v_a\in T_aM$ and $v_a$ the
\emph{geometric representative} of $V_a$.

This canonical association between tangent vectors to $M$ at the
point $a\in M$ and tangent vectors to the fiber $T_aM$ at each one
of its points, establish an isomorphism between fields on $TM$
valued on $TM$, and vertical tangent fields on $TM$. Under this
isomorphism, the field $\dot d$ corresponds to the the vertical
tangent field $\dot D$ on $TM$ such that $\dot D\dot f=\dot f$, for
$f\in\cC^\infty(M)$. This field is the infinitesimal generator for
the group of homotheties of the fibers of $TM$. To avoid any
confusion with the notation for second order differential equations
defined below, we will no longer use the capital notation $\dot D$.

\begin{definition} \textbf{\em (Second Order Differential Equation).}\label{def: second_order_equation}
A vector field $D$ on $TM$ is a \emph{second order differential
equation} when its restriction (as derivation) to the subring
$\cC^\infty(M)$ of $\cC^\infty(TM)$ is $\dot d$.

This is equivalent to have $\pi_*(D_{v_a})=v_a$ at each point
$v_a\in T_aM$ (where $\pi\colon TM\to M$ denotes the canonical
projection).
\end{definition}

\begin{remarkN} \label{obs:diferencia}
\em{The difference between two tangent vector fields on $TM$ which
are second order differential equations, is  a vertical vector
field. Thus the second order differential equations on $TM$ are
sections of an affine bundle modeled on the fiber bundle over $TM$
of the vertical tangent fields. This last one is isomorphic to the
bundle of fields on $TM$ taking values on $TM$.}
\end{remarkN}

\begin{definition}\textbf{\em (Contact System).} \label{def: Pfaff_system}
The \emph{contact system} on $TM$ is the Pfaff system in $TM$ which
consists of all the $1$-forms annihilating all the second order
differential equations. It will be denoted by $\Omega$.
\end{definition}

\begin{remarkN} \label{obs:contacto}
\em{The forms in the contact system also annihilate the
differences of second order differential equations, i.e. all
vertical fields. Therefore, they are horizontal forms; each
$\omega_{v_a}\in\Omega_{v_a}$ is the pull-back to $T^*_{v_a}TM$ of
a form in $T^*_aM$. Now, a horizontal $1$-form kills a second
order differential equation if and only if it kills the field
$\dot d$. Thus the contact system on $TM$ consists of the
horizontal $1$-forms which annihilate $\dot d$.}
\end{remarkN}
\bigskip

\begin{nada}
\label{nada:local0} {\bf Local coordinate expressions.}
\end{nada}
We take local coordinates $(q^1,\dots,q^n)$ in $M$ and
corresponding $(q^1,\dots,q^n,\dot q^1,\dots,\dot q^n)$ in $TM$.
We have, using Einstein summation convention,

\begin{equation*}
\dot d=\dot q^i\frac{\partial}{\partial q^i}\ \ .
\end{equation*}

A vertical field has the expression
 \begin{equation*}
 V=f^i(q,\dot q)\frac{\partial}{\partial \dot q^i} \ \ .
 \end{equation*}

And the one for a second order differential equation is
\begin{equation*}
D=\dot q^i\frac{\partial}{\partial q^i}+f^i(q,\dot
q)\frac{\partial}{\partial \dot q^i}\ \ .
\end{equation*}

Usually we will denote $f^i$ by $\ddot q^i$ understanding that it is
a given function of the $q$'s and $\dot q$'s.

A local system of generators for the contact system $\Omega$, out
of the zero section, is given by
\begin{equation*}
\dot q^idq^j-\dot q^jdq^i\quad (i,j=1,\dots,n)\ .
\end{equation*}

\bigskip

\section{Structure of a second order differential equation relative to a metric} \label{sec:second_order_equations}

Let $T^*M$ be the cotangent bundle of $M$ and $\pi\colon T^*M\to M$
the canonical projection. Recall that the \emph{Liouville form}
$\theta$ on $T^*M$ is defined by $\theta_{\alpha_a}=\pi^*(\alpha_a)$
for $\alpha_a\in T^*_aM$. Abusing the notation we can write
$\theta_{\alpha_a}=\alpha_a$.

The $2$-form $\omega_2=d\theta$ is the natural symplectic form
associated to $T^* M$.

In local coordinates $(q^1,\dots, q^n)$ in $M$, and corresponding
$(q^1,\dots, q^n,p_1,\dots, p_n)$ for $T^*M$, we have

\begin{equation*}
\theta=p_idq^i,\qquad \omega_2=dp_i\wedge dq^i \ .
\end{equation*}

Let $T_2$ be a (non-degenerate) pseudo-Riemannian metric in $M$.
Then we have an isomorphism of vector fiber bundles
\begin{align*}
TM &\to T^*M\\
v_a &\mapsto i_{v_a}T_2
\end{align*}
($i_{v_a}T_2$ is the inner contraction of $v_a$ with $T_2$).

Using the above isomorphism we can transport to $TM$ all
structures on $T^*M$. In particular, we work with the Liouville
form $\theta$ and the symplectic form $\omega_2$ transported in
$TM$ with the same notation.

From the definitions we have for the Liouville form in $TM$, at
each $v_a\in T_aM$,
\begin{equation}\label{Liouville}
\theta_{v_a}=i_{v_a}T_2\ ,
\end{equation}
where the form of the right hand side is to be understood
pulled-back from $M$ to $TM$.

Equation (\ref{Liouville}) is the point expression for:
\begin{equation}\label{Liouville2}
\theta=i_{\dot d}T_2
\end{equation}

On the other hand, from (\ref{Liouville}) we obtain, for each
$\lambda\in\mathbb{R}$:
\begin{equation}\label{Liouville3}
\theta_{\lambda v_a}=\lambda\theta_{v_a}
\end{equation}
(equality as 1-forms in $T^*M$).

Let $V$ denote the vertical field corresponding to $\dot d$; $V$
is the infinitesimal generator of the group of homotheties in the
fibers of $TM$, and from (\ref{Liouville3}) we get
\begin{equation}\label{Liouville4}
L_V\theta=\theta\ ,
\end{equation}
which, by using Cartan's formula (see \cite{KN}, p. 36) and
$i_V\theta=0$ ($\theta$ being horizontal), gives
\begin{equation}\label{Liouville5}
i_Vd\theta=\theta\quad\text{or}\quad i_V\omega_2=\theta\ .
\end{equation}

Taking the values at each $v_a\in T_aM$ and if we put together
(\ref{Liouville}) and (\ref{Liouville5}) it results the following
key lemma.

\begin{lemma} \label{lem:representantevertical}
If $v_a\in T_aM$ and $V_a$ is its vertical representative at each
point of $T_aM$, we have
\begin{equation}\label{vertical}
i_{V_a}\omega_2=i_{v_a}T_2.
\end{equation}
If $V$ is a vertical tangent field on $TM$ and $v$ is the
corresponding field on $TM$ taking values in $TM$, then we have
\begin{equation}\label{vertical2}
 i_V\omega_2=i_v T_2
\end{equation}
 (equality of horizontal forms in $TM$).
\end{lemma}

\begin{definition}\textbf{\em (Kinetic Energy).}\label{cinetica}
The function $T=\frac 12\,\dot\theta$ on $TM$ is the kinetic
energy associated to the metric $T_2$. So, for each $v_a\in TM$,
we have $T(v_a)=\frac 12 \dot\theta(v_a)=\frac 12 T_2(v_a,v_a)$
or, as a function on $TM$, $T=\frac 12 T_2(\dot d,\dot d)$.
\end{definition}

\bigskip

\begin{theorem}\label{teoremaalfa}
The metric $T_2$ establishes a one-to-one correspondence between
second order differential equations and horizontal $1$-forms in
$TM$.

The second order differential equation $D$ and the horizontal
$1$-form $\a$ that correspond to each other are related by
\begin{equation}\label{formulaalpha}
i_D\omega_2+dT+\alpha=0\ .
\end{equation}
\end{theorem}

\begin{proof}
Given a second order differential equation $D$, we define the
$1$-form $\alpha$ by (\ref{formulaalpha}). Now we check that
$\alpha$ is horizontal. For any vertical field $V$ we must prove
that $ \langle \alpha , V \rangle =0$. Using Cartan's formula,
\begin{equation*}
\langle i_D\omega_2 , V\rangle =\langle i_D d\theta , V\rangle =D
\langle \theta , V\rangle -V\langle \theta, D \rangle -\langle
\theta , [D,V] \rangle \ .
\end{equation*}
Now we have $\langle \theta , V\rangle=0$ since $\theta$ is
horizontal. Also we have $ \langle \theta, D
\rangle=\la\theta,\dot d\ra=\dot \theta$, and $ V\dot \theta
=2\langle \theta, v\rangle$, where $v$ is the geometric
representative of $V$ and using that $\dot \theta$ is homogeneous
of the second degree in the variables $\dot q$. Therefore we have
$$
V\langle \theta, D\rangle =2 \langle \theta, v\rangle \ .
$$
Also for  $f \in \cC^\infty (M)$ we have
$$
[D,V] f=-V(Df)=-V \dot f=-v f \ ,
$$
thus the bracket $[D,V]$ is equal to $-v$ up to a vertical field. It
follows that
$$
\langle \theta, [D,V]\rangle =-\langle \theta, v\rangle \ .
$$
And putting all together, we have
$$
\langle i_D\omega_2 , V\rangle =0-2 \langle \theta, v\rangle +
\langle \theta, v\rangle=-\langle \theta, v\rangle \ .
$$
Also
$$
\langle dT , V\rangle =V T=V({\frac 12} \dot \theta)=\langle
\theta , v\rangle \ ,
$$
therefore $\alpha$ is horizontal.

Lemma (\ref{lem:representantevertical}) establishes a linear
isomorphism, $V\mapsto i_V \omega_2$, between vertical vector
fields and horizontal $1$-forms. So, given $\alpha$, adding a
suitable vertical vector field $V$ to a given second order
differential equation $D_0$ we obtain $D=D_0+V$ that corresponds
to $\alpha$ by (\ref{formulaalpha}).
\end{proof}

\begin{definition}\textbf{\em (Geodesic Field).} \label{campogeodesico}
The geodesic field of the metric $T_2$ is the second order
differential equation, $D_G$, corresponding to $\alpha=0$:
\begin{equation}\label{ecuaciongeodesica}
i_{D_G}\omega_2+dT=0 \ .
\end{equation}
\end{definition}

The projection to $M$ of the curves solution of $D_G$ in $TM$ are
the geodesics of $T_2$.

The geodesic field $D_G$ is chosen as the origin in the affine
bundle of second order differential equations. With this choice we
establish a one-to-one correspondence between second order
differential equations and vertical tangent fields.
$$
D\longleftrightarrow D-D_G=V.
$$
And recalling that to a vertical field $V$ there canonically
corresponds a field $v$ on $TM$ taking values in $TM$ (the geometric
representative of $V$), we define:

\begin{definition}\textbf{\em (Covariant Value).} \label{covariante}
We define the covariant value of the second order differential
equation $D$, denoted by $D^\nabla$, as the field in $TM$ taking
values in $TM$ corresponding canonically to $D-D_G$.
\end{definition}

\begin{remarkN}
\em{ The covariant value $D^\nabla$ yields at each point $v_a\in
TM$ the acceleration at the point $a\in M$ of a particle with a
trajectory solution of $D$ and having $v_a$ as tangent vector at
the point $a\in M$:
$$
D^\nabla_{v_a}=\nabla_{v_a} v \ ,
$$
where $v$ is the vector field tangent along the trajectory.}
\end{remarkN}

From this definition, Lemma (\ref{lem:representantevertical}), and
the definition of $D_G$ it is straightforward to prove:

\begin{theorem}\label{teoremacovariante}
The horizontal $1$-form corresponding to the second order
differential equation $D$ is related to its covariant value by
\begin{equation}\label{relacionformacovariante}
i_{D^\nabla}T_2=-\alpha\ .
\end{equation}
\end{theorem}

\begin{definition}\label{alfagradiente}
Given a horizontal 1-form $\a$, $\textrm{grad}\,\a$ is the field
on $TM$ taking values in $TM$ such that
\begin{equation}\label{alfagradiente1}
i_{\textrm{grad}\,\a}T_2=\a\ .
\end{equation}

In particular, when $\a=dU$ for $U\in\cC^\infty(M)$,
$\textrm{grad}\,\a$ equals $\textrm{grad}\,U$.

So, Theorem (\ref{teoremacovariante}) gives
\begin{equation}\label{alfagradiente2}
D^\nabla=-\textrm{grad}\,\a\ .
\end{equation}
\end{definition}

\begin{remarkN} \label{remark:covariant}
\em{When $T_2$ is a second order covariant symmetric tensor on $M$,
we can define on $TM$ a ``Liouville form'' $\theta$ by
(\ref{Liouville}), and a closed 2-form $\o_2=d\theta$. Fomulae
(\ref{Liouville2}), (\ref{Liouville3}), (\ref{Liouville4}),
(\ref{Liouville5}) remain valid, as also Lemma
(\ref{lem:representantevertical}). In Theorem (\ref{teoremaalfa}) is
still true that (\ref{formulaalpha}) assigns a horizontal 1-form
$\a$ to each second order differential equation $D$, although not in
a one to one way, as a rule.}
\end{remarkN}

\begin{nada}
\label{nada:local1} {\bf Local coordinate expressions.}
\end{nada}

Consider an open set of $M$ with coordinates $q^1,\dots,q^n$ and the
corresponding open set in $TM$ with coordinates $q^1,\dots,q^n,\dot
q^1,\dots,\dot q^n$.

If the expression in local coordinates of $T_2$ is
\begin{equation}\label{metrica}
T_2=g_{jk}(q)\ dq^jdq^k
\end{equation}
then the local equations for the isomorphism $TM\approx T^*M$ are
\begin{equation}\label{isomorfismo}
p_j=g_{jk}(q)\ \dot q^k.
\end{equation}

The Liouville form in $TM$ is given by
\begin{equation}\label{Liouvilletangente}
\theta=g_{jk}(q)\ \dot q^kdq^j
\end{equation}
and the symplectic form in $TM$ by
\begin{equation}\label{simplectica}
\omega_2=g_{jk} d\dot q^j\wedge dq^k+
    \frac{\partial g_{jk}}{\partial q^l}\dot q^jdq^l\wedge dq^k\ .
\end{equation}

For the kinetic energy we have, locally,
\begin{equation}\label{energiacinetica}
 T=\frac 12\, g_{ij}\dot q^i \dot q^j,
\end{equation}
so that
\begin{equation}\label{p}
p_j=\frac {\partial T}{\partial \dot q^j} \ .
\end{equation}

Let the second order differential equation $D$ be given by
\begin{equation}\label{formulaD}
D=\dot q^i\frac{\partial}{\partial q^i}+\ddot q^i
\frac{\partial}{\partial \dot q^i} \ ,
\end{equation}
where the $\ddot q^i$'s are given function of $q$'s and $\dot q$'s.
The local expression (\ref{simplectica}) for $\omega_2$ gives
$$
i_D\omega_2=-g_{ij} \dot q^j d\dot q^i+ g_{ij}\ddot q^i dq^j+
\frac{\partial g_{ij}}{\partial q^k} \dot q^i(\dot q^k dq^j-\dot q^j
dq^k) \ .
$$
On the other hand, we have
$$
dT=g_{ij}\dot q^i d\dot q^j+ \frac 12\,\frac{\partial
g_{ij}}{\partial q^k} \dot q^i\dot q^j dq^k \ .
$$
Replacing this equality into the precedent expression we get
 \begin{align*}
 i_D\omega_2 &= -dT+
         \frac 12\,\frac{\partial g_{ij}}{\partial q^k}\dot q^i\dot q^j dq^k
        +g_{ij}\ddot q^i dq^j
        +\frac{\partial g_{ij}}{\partial q^k}
              \dot q^i(\dot q^k dq^j-\dot q^j dq^k)\\
     &=-dT+ g_{ij}\ddot q^i dq^j
      +\frac 12\left(
            \frac{\partial g_{ik}}{\partial q^j}
            +\frac{\partial g_{jk}}{\partial q^i}
            -\frac{\partial g_{ij}}{\partial q^k}
            \right) \dot q^i\dot q^j dq^k\\
     &=-dT+(g_{ik}\ddot q^i+\Gamma_{ij,k}\dot q^i\dot q^j) dq^k \ ,
  \end{align*}
where the $\Gamma$'s are the Christoffel symbols of $T_2$.

Now
\begin{equation}\label{alfa}
\alpha=-g_{lk}(\ddot q^l+\Gamma_{ij}^l\dot q^i\dot q^j) dq^k
\end{equation}
is the horizontal $1$-form related to $D$ by formula
(\ref{formulaalpha}).

For the geodesic field we have,

\begin{equation}\label{geodesico}
D_G=\dot q^i\frac{\partial}{\partial q^i}-\Gamma_{ij}^l \dot q^i
\dot q^j \frac{\partial}{\partial \dot q^l}
\end{equation}
and, finally, the covariant value of $D$ is
\begin{equation}\label{coordenadascovariante}
 D^\nabla=(\ddot q^l+\Gamma_{ij}^l\dot q^i\dot q^j)
              \frac{\partial}{\partial q^l}\ .
\end{equation}
\bigskip

\section{Newton-Lagrange mechanics of a free system} \label{sec:Newton-Lagrange}

\begin{definition} \textbf{\em (Mechanical System).} A \emph{mechanical system} $(M,T_2, \alpha )$
is a manifold  $M$ (the \emph{configuration space}) provided with a
pseudo-Riemannian metric $T_2$ (to which corresponds a \emph{kinetic
energy} $T$ by (\ref{cinetica})), and a 1-form $\alpha$ on $TM$ (the
\emph{work-form} or \emph{force form}).
\end{definition}

\begin{postulate}\textbf{\em (Newton Law).}\label{Newton}
The tangent field $D$ on $TM$ such that
\begin{equation}\label{eq:Newton}
i_D\omega_2+dT+\alpha=0
\end{equation}
is a second order differential equation.
\end{postulate}

From Theorem (\ref{teoremaalfa}), we have
\begin{theorem}\label{teorematrabajo}
The force-form $\a$ of a mechanical system obeying Newton's law is
horizontal.
\end{theorem}

\begin{definition}\textbf{\em (Work).}\label{trabajo}
A \emph{trajectory} of the mechanical system is a curve in $TM$
solution of $D$. The integral of $\a$ along a given trajectory is
called \emph{work} done by the system.
\end{definition}

 Contracting with $D$ in (\ref{eq:Newton}), we get
 \begin{equation}
 DT+i_D\alpha=0,
 \end{equation}
 and, by integrating along a trajectory $c$,
 \begin{equation}
 \int_c(dT+\alpha)=0
 \end{equation}
 which is an expression for the \emph{law of conservation of energy}:
 the work done by the system equals the loss of kinetic energy.

\begin{definition}\textbf{\em (Conservative System).}
When $\a$ is an exact differential form, the mechanical system
$(M,T_2,\alpha)$ is called \emph{conservative}. The function $U$
(determined up to an additive constant) such that $\a= dU$ is
called the \emph{potential energy} of the system.
\end{definition}

\begin{theorem}
In a conservative system $(M,T_2, dU)$, the potential energy $U$
is a function on $M$ (lifted to $TM$).
\end{theorem}

 \begin{proof}
 $dU$ is horizontal.
 \end{proof}

 \begin{definition}\textbf{\em (Hamiltonian).}
  In a conservative system $(M,T_2,dU)$, the function $H=T+U$ in $TM$ is called
  the \emph{total energy} or \emph{hamiltonian of the system}.
 \end{definition}

 Equation (\ref{eq:Newton}) for a conservative system is
 \begin{equation}\label{hamiltoniano}
 i_D\omega_2+dH=0 \ ,
 \end{equation}
 which, read in $T^*M$, is the set of \emph{Hamilton's canonical
 equations}.

 Equation (\ref{alfagradiente2})  corresponds to the classical Newton's Law
 $\overrightarrow{F}=m\cdot \overrightarrow{a}$. For a
 conservative system:
 \begin{equation}\label{gradiente}
 D^\nabla=-\textrm{grad}\,U\ .
 \end{equation}
or, in local coordinates,
 \begin{equation}
 \frac{d^2 q^l}{dt^2}
           +\Gamma_{ij}^l\frac{d q^i}{dt}\frac{d q^j}{dt}
     =-g^{lh}\frac{\partial U}{\partial q^h}\ .
 \end{equation}

 For the geodesic field ($\alpha=0$), we obtain the equations of \emph{geodesics}
 for $T_2$
 \begin{equation}
 \frac{d^2 q^l}{dt^2}
           +\Gamma_{ij}^l\frac{d q^i}{dt}\frac{d q^j}{dt}
     =0
 \end{equation}
as a generalization of the classical Euler's theorem (see
\cite{Euler}) for the movement of a point on a surface in absence
of external forces.

 For a general 1-form of force $\a$, the corresponding field $D$
 differs of the geodesic one, $D_G$, by the vertical field
 $D-D_G$; it is natural to consider this field as the ``cause''
 bringing the system to move out the geodesics. So, we give the
 following
 \begin{definition}\textbf{\em (Force).} \label{force}
 A \emph{force} on $M$ is a vertical field $V$ on $TM$. The field
 $v$ on $TM$ which takes values in $TM$ corresponding canonically to
 $V$, will be called the \emph{geometric expression} of $V$, or
 the \emph{geometric representative} of $V$. The force associated
 to a second order differential equation $D$ by the metric $T_2$
 is $D-D_G$. Its geometric expression is $D^\nabla$.
 \end{definition}

 \section{Constrained systems}\label{sec:constrained}

 \begin{definition}\textbf{\em (Constraints).}
 A \emph{constrained mechanical system} is a mechanical system
 $(M,T_2,\alpha)$ together with a Pfaff system $\L$ on $TM$. Each 1-form in $\L$
 is a \emph{constraint}.
 \end{definition}

 \begin{postulate}\textbf{\em (Newton-Lagrange Law).}\label{Newtonligaduras}
 Given the constrained mechanical system $(M,T_2,\alpha,\L)$, there exists
 a tangent vector field $\overline D$ on $TM$, which is a second order
 differential equation, and satisfies the congruence
 \begin{equation}\label{eq:Newtonligaduras}
 i_{\overline D}\omega_2+dT+\alpha\equiv 0\, \textrm{mod}\,
 \Lambda
 \end{equation}
 and also the \emph{principle of virtual works} (stated below
 (\ref{principle})).
 \end{postulate}

  By Theorem (\ref{teorematrabajo}) $\a$ is horizontal. Since $\overline D$
is a second order differential equation, the $\b$'s in $\L$
transforming (\ref{eq:Newtonligaduras}) into an equality, have to be
horizontal. Thus, we can always replace $\L$ by its intersection
with the space of horizontal forms on $TM$. In the sequel, \emph{we
will assume that $\L$ is a Pfaff system of horizontal forms on
$TM$}. Therefore, for each $\b$ in $\L$, it makes sense to consider
the function $\dot\b$ on $TM$.

 \begin{definition}\textbf{\em (Admissible
 State.)}\label{admisible}
 An \emph{admissible state} for the constrained mechanical system
 $(M,T_2,\a,\L)$ is a point $v_a\in TM$ such that $\dot\b
 (v_a)=0$ for each $\b$ in $\L$. The set consisting of all admissible
 states is denoted by $\cL$.
 \end{definition}

  \begin{principleA}\label{principle}
  A curve in $TM$ which is a  solution of $\overline D$ and passes
  through a point $v_a\in\cL$, remains entirely in $\cL$.
  \end{principleA}

  When $\cL$ is a submanifold of $TM$ this is equivalent to
  $\overline D$ being tangent to $\cL$.

  \begin{remarkN}
  \em{Congruence (\ref{eq:Newtonligaduras}) is a form of D'Alembert
  principle of equilibrium between applied, inertial, and constraint
  forces. Definition (\ref{admisible}) selects as admissible
  velocities those for which, in the corresponding ``infinitesimal
  displacements'', the constraint forces do not work. The principle
  (\ref{principle}) means that the system remains in admissible
  states: the constraint forces never work. (See, e.g.
  \cite{Sommerfeld}, Sect.10).}
  \end{remarkN}

 \begin{nada} \textbf{\em Local computation for $\overline D$.}
 \end{nada}
 We take local coordinates in  $M$ and a local basis
 $\{\beta_1,\dots,\beta_r\}$ for $\Lambda$ in $TM$:
 \begin{equation}
 \beta_k=B_{kj}(q^1,\dots,q^n,\dot q^1,\dots,\dot q^n)\, dq^j\qquad
 (k=1,\dots,r)
 \end{equation}

  Each $\beta_k$ corresponds to a vertical tangent field $V_k$
  (a \emph{constraint force}) by
  \begin{equation}\label{eq:beta}
 i_{V_k}\omega_2=\beta_k
 \end{equation}

  Let $D$ be the field for the free system $(M,T_2,\alpha)$: $$i_D\omega_2 + dT + \alpha=0\ .$$
  Then, the congruence (\ref{eq:Newtonligaduras}) can be written as:
 \begin{equation}\label{Dligado}
 \overline D=D+\lambda^1V_1+\cdots+\lambda^rV_r\ ,
 \end{equation}
 where the $\lambda^k$'s are certain (local) functions on $TM$, called
 \emph{Lagrange multipliers}.
 Our problem is to compute the $\l$'s such that $\overline
 D$ satisfies the principle of virtual works (\ref{principle}).

 \emph{We assume that} $\cL$ \emph{is a submanifold of}
 $TM$. In this situation we compute the $\l$'s by requiring $\overline D$ to
 be tangent to $\cL$.

  We have the local expressions
 \begin{equation}\label{eq:fuerzasligadura}
 V_k=a_k^h(q^1,\dots,q^n,\dot q^1,\dots,\dot q^n)
           \frac{\partial}{\partial\dot q^h}\qquad (k=1,\dots,r),
 \end{equation}

 The local equations of $\cL$ are
 $\dot\beta_1=0,\dots,\dot\beta_r=0$. Then, the conditions for $\overline D$
 being tangent to $\cL$ are:
 \begin{equation}\label{eq:multiplicadores}
 \lambda^k a_k^h
    \frac{\partial(B_{lj}\dot q^j)}{\partial\dot q^h}
     \equiv -D(B_{lj}\dot q^j)\,\,\textrm{mod}\,
     (\dot\beta_1,\dots,\dot\beta_r)
 \end{equation}

 Using coordinates, (\ref{eq:beta}) is:
 \begin{equation}
 g_{jh} a_k^h=B_{kj}
 \end{equation}
 and (\ref{eq:multiplicadores}) becomes:
 \begin{equation}\label{virtual}
 \lambda^k a_k^h
    \frac{\partial(g_{js}a_l^s\dot q^j)}{\partial\dot q^h}
     \equiv -D(g_{jh}a_l^h\dot q^j)\,\,\textrm{mod}\,
     (\dot\beta_1,\dots,\dot\beta_r)\ .
 \end{equation}

 The discussion of (\ref{virtual}) is carried out according to the supplementary
 hypothesis in each particular case.
 We start with the most common one:

 \begin{definition}\textbf{\em (Linear Constraints).}\label{lineal}
 $\L$ is a system of \emph{linear constraints} when it is generated, as a Pfaff system on
 $TM$, by a Pfaff system $\L_M$ on $M$ (lifted to $TM$).
 \end{definition}

 Let $r$ be the rank of $\L_M$. For each $a\in M$, the vectors
 $v_a\in T_aM$ annihilated by the forms in $(\L_M)_a$ form
 a $(n-r)$-dimensional subspace  $\cL_a\subset T_aM$. The
 collection of all $\cL_a$, when $a$ varies in $M$, is the set
 $\cL$ of admisible states. Therefore, in the case of linear constraints,
 $\cL$ is the vector subbundle of $TM$ corresponding to the
 distribution of vector fields incident with (i.e. annihilated by) $\L_M$.

 In this case, the $a_k^h$ in (\ref{eq:fuerzasligadura}) can be chosen free from the
 $\dot q$'s, and the left member  in (\ref{virtual}) becomes

 \begin{equation}\label{virtual3}
 \lambda^k a_k^h g_{js}a_l^s\delta_{jh}
  =\lambda^k a_k^h g_{hs}a_l^s
  =\lambda^k \langle v_k, v_l\rangle\ ,
 \end{equation}
 where  $v_k$, $v_l$ are the geometric representatives of $V_k$, $V_l$
 (see Definition (\ref{force})), and $\langle\, ,\, \rangle$ is
 the scalar product with respect to $T_2$.

 When $T_2$ is positive definite, the matrix with entries
 $\langle v_k, v_l\rangle$ is non degenerate, because the $v$'s are linearly independent.
 It follows that (\ref{virtual}) can be solved when we replace
 congruence by equality giving unique solutions for the $\l$'s by:
 \begin{equation}\label{virtual4}
 \lambda^k \langle v_k, v_l\rangle=-D\dot\beta_l\ .
 \end{equation}

 With these values for the $\l$'s, the $\overline D$ in
 (\ref{Dligado}) gives:
 \begin{equation}\label{virtual2}
 \overline D\dot\beta_l=D\dot\beta_l+\lambda^kV_k\dot\beta_l
           =D\dot\beta_l+\lambda^k\langle\beta_l, v_k\rangle
           =D\dot\beta_l+\lambda^k\langle v_l,v_k\rangle
           =0
 \end{equation}

 In the open set of $TM$ where we work, the field
 $\overline D$ satisfies the conditions imposed by
 (\ref{Newtonligaduras}). Also $\overline D$ depends upon the basis
 $\{\b_1,\dots,\b_r\}$, but its restriction to $\cL$ does not
 depend.

 The local fields $\overline D$ can be used to build, by means of
 an appropriate partition of unity, a global field $\overline D$
 on $TM$ satisfying (\ref{principle}). The uniqueness of
 $\overline D\mid_\cL$ follows from the local uniqueness.

 We have proved:
  \begin{theorem}\label{teoremaNewtonligaduras}
  When  $T_2$  is positive definite  and $\L$ is a linear system of constraints,
  then there exists
  on $TM$ a second order differential equation $\overline D$ that satisfies the
  Newton-Lagrange law (\ref{Newtonligaduras}).
  The restriction of $\overline D$ as tangent field to the bundle
  $\cL$ of admisible states is fixed by the condition settled in
     (\ref{principle}).
  \end{theorem}

  \begin{remarkN}\textbf{\em (About non linear constraints).}
\em{In the general case of a Pfaff system $\L$ of horizontal forms
on $TM$, and assuming $T_2$ to be positive definite, the left member
of (\ref{virtual}) restricted to the $0$ section of $TM$ is
  (\ref{virtual3}). Thus, in a neighborhood of the $0$ section, there
  exists a unique solution for the $\l$'s when in (\ref{virtual}) we replace
  $\equiv$ by $=$. And, in the analytic case, there exists  a global
  $\overline D$, with possible singularities out of the $0$ section (i.e.
  ``for large velocities'').}
  \end{remarkN}

  The most important case of linear constraints is that of
  holonomic constraints:
 \begin{definition}\textbf{\em (Holonomic constraints).}
 A system $\L$ of constraints is called \emph{holonomic} when it
 is linear and generated by a completely integrable Pfaff system
 $\L_M$ on $M$.
 \end{definition}

 In this case, let $\{dB_1,\dots,dB_r\}$ be a local basis for $\L_M$. The
 local equations for $\L$ are  $\dot B_1=0,\dots,\dot B_r=0$, and
 (locally) $\L$ is the union of the tangent bundles to the submanifolds of
 $M$ given by equations $B_1=b_1$, $\dots$, $B_r=b_r$ ($b_i$ constants). Let
 $N\subset M$ be one of these submanifolds. In the embedding
 $N\subset M$, the metric $T_2$ on $M$ specializes as $T_{2N}$.
 In the embedding $TN\subset TM$, the Liouville form $\theta$ on $TM$,
 specializes as the Liouville form on $TN$
 corresponding to the metric $T_{2N}$ (this is a consequence of
 (\ref{Liouville})). Then, $\omega_2$ in $TM$ specializes as $\omega_{2N}$ in $TN$,
 which is also the symplectic form corresponding to $T_{2N}$.

 The field $\overline D$ on $TM$ corresponding to the mechanical system
 $(M,T_2,\a,\L)$ is tangent to $TN$, because the local
 equations for  $TN$ in $TM$ are:
 $$B_1=b_1,\dots, B_r=b_r,\dot B_1=0,\dots, \dot
 B_r=0,$$
 and, for each $v_a\in T_aN$ we have
 $\overline D_{v_a}(B_k)=v_a(B_k)=0$ and
  $\overline D_{v_a}\dot B_k=0$, by (\ref{principle}).

 Now, in the Newton-Lagrange equation
 $$i_{\overline D}\omega_2+dT+\alpha=\sum_{k=1}^r\lambda^k\,
 dB_k$$
 everything specializes to $TN$, and with such a specialization, we obtain
 $$i_{\overline D_N}\omega_{2N}+dT_N+\alpha_N=0.$$

 \begin{theorem} When the constraints are holonomic, the field $\overline D$
 corresponding to the system $(M,T_2,\a,\L)$, tangent to the
 submanifold $\cL$ of admisible states, is also tangent to the submanifolds
 $TN\subset\cL$, for each solution $N\subset M$ of the Pfaff system $\L_M$.
 The restriction of $\overline D$ to $TN$ is the field corresponding to the
 free system $(N, T_{2N},\a_N)$.
 \end{theorem}
 \bigskip

 \begin{nada}\textbf{\em Analysis of general linear constraints.}
 \label{generallineal}
 \end{nada}
 Let $D$ be a second order differential equation, $\b$ a
 1-form on $M$. Working in local coordinates we have
 $$\beta=B_l(q^1,\dots,q^n)\, dq^l\qquad
                \dot\beta=B_l(q^1,\dots,q^n)\, \dot q^l,$$
 and
 \begin{align*}
 D\dot\beta &=B_l\ddot q^l+D(B_l)\, \dot q^l=
                 \langle\beta,
                    D^\nabla-\Gamma_{hk}^l\dot q^h\dot q^k
                      \frac{\partial}{\partial q^l}\rangle
                       +\dot d(B_l)\dot q^l\\
            &=\langle\beta, D^\nabla\rangle+
                 \dot q^h\dot q^k
                   \left(\frac{\partial B_h}{\partial q^k}
                          -B_l\Gamma_{hk}^l\right)\\
            &=\langle\beta, D^\nabla\rangle+
                    \langle \nabla_{\dot d}\beta,\dot d\rangle\ ,
 \end{align*}
where $\nabla_{\dot d}\beta$ is an horizontal 1-form on $TM$ and can
be paired by duality with $\dot d$.

 For the geodesic field, we obtain
 $$D_G\dot\b=\langle \nabla_{\dot d}\beta,\dot d\rangle\ .$$

 Let $v$ be the field such that $i_vT_2=\beta$.
 We have
 $$\langle \nabla_{\dot d}\beta,\dot d\rangle=
      \langle \nabla_{\dot d}(i_vT_2),\dot d\rangle=
       T_2(\nabla_{\dot d}v,\dot d)=
          \textrm{II}_v(\dot d,\dot d),$$
 where $\textrm{II}_v$ is the \emph{second fundamental form} associated to
 the field $v$.

 Putting all together, we obtain
 \begin{equation}\label{eq:segundaforma}
 D\dot\beta=\langle\beta, D^\nabla\rangle+D_G\dot\beta=
           \langle\beta, D^\nabla\rangle+
                \textrm{II}_v(\dot d,\dot d).
 \end{equation}

  This formula will be used to clarify the behaviour of a
  mechanical system with linear constraints.

  Let $(M,T_2,\alpha,\Lambda)$ be such a system, let
 $\{\beta_1,\dots,\beta_r\}$ be a local basis for $\L_M$, and
 $v_k=\textrm{grad}\,\b_k$, $V_k$ the vertical representative of
 $v_k$, so that $i_{V_k}\omega_2=\beta_k$ (Lemma (\ref{lem:representantevertical})).
 We write $II_k$ for the second fundamental form corresponding to
 $v_k$.

 By a suitable choice of the basis, we set
 $$\langle v_h,v_k\rangle=\delta_{hk},\qquad (h,k=1,\dots,r).$$

 Then, equations (\ref{virtual4}) for the Lagrange multipliers
 are
 $$\l^l=-D\dot\b_l,$$
 when $D$ is the field for the free system $(M,T_2,\a)$. And the
 field $\overline D$ for the constrained system is, according
 (\ref{Dligado})
 \begin{equation}\label{eq:Dortonormal}
 \overline D=D-\sum_{k=1}^r(D\dot\beta_k)V_k,
 \end{equation}

 From   (\ref{eq:segundaforma}) and
 (\ref{eq:Dortonormal}), we obtain  for the covariant  values,
 \begin{equation}\label{valorescovariantes}
 \overline D^\nabla=
  D^\nabla-\sum_{k=1}^r(D\dot\beta_k)v_k=
     D^\nabla-\sum_{k=1}^r\langle v_k,D^\nabla\rangle v_k
          -\sum_{k=1}^r \textrm{II}_k(\dot d,\dot d)v_k
  \end{equation}

  The sum $D^\nabla-\sum_{k=1}^m\langle v_k,D^\nabla\rangle v_k$
  is the orthogonal projection of $D^\nabla$ on the distribution
  $\cL$ (in this case, the orthogonal distribution to
  $v_1,\dots,v_r$). The sum
  $-\sum_{k=1}^r \textrm{II}_k(\dot d,\dot d)v_k$ is the only term
  surviving when $D$ is the geodesic field ($\a=0$); then, it
  is $\overline D_G^\nabla$, the field corresponding to the
  constrained system $(M,T_2,0,\L)$.

  The formula
  $\overline D_G^\nabla=-\sum_{k=1}^r \textrm{II}_k(\dot d,\dot
  d)v_k$,
  in the particular case of holonomic constraints, express a well
  known geometric theorem: when $N$ is a submanifold of $M$, the
  acceleration in $M$ of a geodesic of $N$ is orthogonal to $N$.

  Formula (\ref{valorescovariantes}) is a related to classical results;
  see e.g. Prange \cite{Prange}, pag. 558.
  \bigskip

  \section{The class of time and the calculus of variations}

  As usual, differential 1-forms on $M$ are also considered as 1-forms on $TM$,
  by means of the natural pull-back.
  A 1-form $\a$ on $M$ defines the function $\dot\a$ on $TM$ and,
  in the open set where $\dot\alpha$ does not vanish, it defines the
  1-form $\frac{\alpha}{\dot\alpha}$.
  For each second order  differential equation $D$ we have
  $\la \frac{\alpha}{\dot\alpha}, D\ra=1$. So, for two 1-forms
  $\a$, $\b$ on $M$ we have
  $\la \frac{\a}{\dot\a}-\frac{\b}{\dot\b}, D\ra=0$ in the open
  set of $TM$ where neither $\dot\a$ nor $\dot\b$ vanish. From the
  definition of the contact system (\ref{def: Pfaff_system}) it
  follows that we have in that open set
    $$\frac{\alpha}{\dot\alpha}\equiv
  \frac{\beta}{\dot\beta}\,\,\,\textrm{mod}\,\Omega\ .$$

  \begin{definition}\textbf{\em (Class of
  Time).}\label{classoftime}
  Let $U$ be an open set of $TM$ non intersecting the zero section.
  A 1-form  $\tau$ defined on  $U$ belong to the \emph{class of
  time} when at each point $v_a\in U$ it satisfies
   $$\t_{v_a}\equiv\frac{\a_{v_a}}{\dot
   \a(v_a)}\,\,\emph{\textrm{mod}}\,\O_{v_a}$$
   for each 1-form $\a$ on $M$ such that $\dot\a(v_a)\ne 0$.
  \end{definition}

  The 1-forms belonging to the class of time are horizontal.
  Therefore we can apply them to the field $\dot d$. The
  following proposition is obvious
  \begin{proposition} \label{tiempo}
  A 1-form $\t$ on an open set $U\subset TM$ non intersecting the zero section belongs
  to the class of time if and only if it is horizontal and satisfies $\la\t,\dot
  d\ra=1$. Therefore $\la \t, D\ra$=1 for each second
  order differential equation $D$.
  \end{proposition}

  \begin{definition}\textbf{\em (Infinitesimal contact
  transformation).}\label{infinitesimal}
  A tangent field $\delta$ on $TM$ is  a \emph{infinitesimal
  contact transformation} if  $L_\delta\Omega\subseteq\Omega$ (i.e., for each 1-form
  $\s$ in $\O$, $L_\d\s$ is in $\O$).
  \end{definition}

  When $\d$ generates an uniparametric group of automorphisms of
  $TM$, the condition $L_\delta\Omega\subseteq\Omega$ means that
  this group transforms solutions of $\O$ into solutions of $\O$,
  i.e., maps into each other curves which are tangent to
  second order differential equations or vertical fields.

  \begin{theorem} \label{variacion}
  Let $\d$ be an infinitesimal contact transformation on $TM$,
  which is projectable to $M$ as a tangent field. The
  following two properties are equivalent:
  \begin{enumerate}
  \item \label{variacion1}
   $\delta$ leaves invariant the class of time: for each 1-form
  $\a$ on $M$, in the open set of $TM$ where $\dot\a\ne 0$, we
  have:
  $$L_\delta\left(\frac{\a}{\dot
  \a}\right)\equiv 0\,\,\,{\text{\emph{mod}}}\,\, \Omega\ .$$
  \item\label{variacion2}
   $\delta$ commutes with $\dot d$
    $$\dot d\circ\delta =\delta\circ\dot d$$
  as derivations from $\cC^\infty(M)$ to $\cC^\infty(TM)$.
  \end{enumerate}
  (Note that $\dot d\circ\delta$ does make sense because
  $\d$ is projectable to $M$, thus maps $\cC^\infty(M)$ into $\cC^\infty(M)$).
  \end{theorem}

  \begin{proof}
  Let $D$ be any second order differential equation and let $\a$ be a
  1-form on $M$. In the open set of $TM$ where $\dot\a\ne 0$, we
  have
  $$0=\d(1)=\d\la\frac{\a}{\dot\a},D\ra
           =\la L_\d\left(\frac{\a}{\dot\a}\right),D\ra
            +\la \frac{\a}{\dot\a},[\d,D]\ra\ .$$

            The first term in the last sum is $0$ for arbitrary
  $D$ if and only if $\d$ satisfies (\ref{variacion1}). The second term is $0$ for
  arbitrary $\a$ if and only if $[\d,D]$ is vertical, and this is
  equivalent to
  (\ref{variacion2}).
  \end{proof}

  \begin{definition}\textbf{\em (Infinitesimal Variation). }
  An infinitesimal contact transformation that is projectable to
  $M$ and satisfies the equivalent conditions of Theorem
  (\ref{variacion}) is called an \emph{infinitesimal variation}.
  \end{definition}

  \begin{theorem}
  For each vector field $v$ on $M$, there exists a unique
  infinitesimal variation $\delta_v$ on $TM$ which projects to $M$
  as $v$.
  \end{theorem}
  \begin{proof} When $\d_v$ exists, (\ref{variacion2}) of
  (\ref{variacion}) gives, for each $f\in\cC^\infty(M)$,
  $$\d_v\dot f=\dot d(vf)\ .$$

  Thus, $\d_v$ is uniquely determined by $v$.

  Given a coordinate system $q^1,\dots,q^n$ on an open set
  $U\subseteq M$, let
  $v=a^i(q^1,\dots,q^n)\frac{\partial}{\partial q^i}$ be the local
  expression for $v$. Defining
  $\d_v=a^i\frac{\partial}{\partial q^i}
    + \dot a^i(q^1,\dots,q^n)\frac{\partial}{\partial \dot q^i}$ we
    check that $\d_v$ is an infinitesimal contact transformation
    by computing the Lie derivatives of
    the forms $\dot q^i dq^j-\dot q^j dq^i$ (local generators of $\O$ on
    $TM$ out of the zero section). Then, from its actual definition, $\d_v$ fulfills (\ref{variacion2}) of
    (\ref{variacion}) in $U$. The uniqueness shows that the local $\d_v$'s
    patch together  giving a $\d_v$ globally defined on the whole of $TM$.
    \end{proof}

  \begin{definition}\textbf{\em (Prolongation of a Field).}
  $\delta_v$ is called the \emph{prolongation of} $v$ \emph{to} $TM$.
  \end{definition}

 \begin{remarkN}
 \em{
 The tangent bundle $TM$ is the space of contact elements $M_1^1$
 in the sense of Weil \cite{Weil}. Formula (\ref{variacion2}) of
 (\ref{variacion}) means that $\d_v$ is the canonical
 prolongation of $v$ from $M$ to $M_1^1$. Also, $\d_v$ is the
 infinitesimal generator of the prolongation to $TM$ of the group
 (or local group) of automorphisms of $M$ generated by $v$.}
 \end{remarkN}

   \begin{remarkN}
   \em{
  (\ref{variacion2}) of (\ref{variacion}) is the
  `$d\circ\delta=\delta\circ d$' in the classical texts of
  Mechanics; e.g. in Sommerfeld
  \cite{Sommerfeld}, formulae (9) and (9a) of page 175.
  Our formula (\ref{variacion1}) of (\ref{variacion}) is reminiscent of
  ``keeping fixed the time'' in the reasoning from Sommerfeld in
  pages 175, 176. Sommerfeld attributes to Euler the formula
  $d\d=\d d$. For this reason, we shall call (\ref{variacion2}) of
  (\ref{variacion})
  \emph{Euler's commutation formula}.}
  \end{remarkN}

  The starting point for the applications of variational methods in
  Mechanics is the following theorem (see Prange
  \cite{Prange}, ``Zentralgleichung von Lagrange'', in page 531).

  \begin{theorem} \label{zentral}
  Let $M$ be a manifold provided with a pseudo-Riemannian  metric
  $T_2$, $\theta$ the corresponding Liouville form on $TM$, $T$
  the kinetic energy. For any second order differential equation
  $D$ and any infinitesimal variation $\d$, we have
  \begin{equation}\label{eq:zentral}
  D\langle\theta, \d\rangle=
    \delta T-\langle\alpha, \delta\rangle\ ,
  \end{equation}
   where  $\a$ is the work form corresponding to $D$ by
    Theorem (\ref{teoremaalfa}).
    \end{theorem}
  \begin{proof}
  The Cartan formula for the Lie derivative, with
  Definition (\ref{cinetica}) for $T$ and Theorem
  (\ref{teoremaalfa})
  gives
  $$
  L_D\theta =i_D d\theta+d\langle\theta, D\rangle
             =i_D\omega_2+2 dT=-dT-\alpha+2dT=dT-\alpha\ ,$$
  thus,
  $$D\langle\theta, \delta\rangle=\langle dT-\alpha, \delta\rangle
            +\langle\theta, [D,\delta]\rangle
            =\d T-\la\a,\d\ra\ ,$$
  because $[D,\delta]$  is vertical as follows from Euler's
  commutation formula.
  \end{proof}

 \begin{definition}\textbf{\em (Lagrangian function).}
 In a conservative system $(M,T_2,dU)$, the function $L=T-U$ is called
  \emph{the Lagrangian function} of the system.
 \end{definition}

 Applying Theorem (\ref{zentral}) to a conservative system, we obtain:

 \begin{theorem}\textbf{\em (Hamilton's Principle).}\label{hamilton}
 Let $(M,T_2,dU)$ be a conservative mechanical system, $D$ the corresponding
 second order differential equation and $L$ the lagrangian function. For each
 infinitesimal variation $\delta$, we have
 \begin{equation}\label{eq:hamilton}
 D\langle \theta, \delta\rangle=\delta L\ .
 \end{equation}
 \end{theorem}

  The classical integral version of Hamilton's principle follows from
  (\ref{hamilton})
 as we show now.

 Let $c\colon[t_0,t_1]\to M$ be a parameterized curve such that
 its canonical lift to $TM$ is a solution of $D$. Let $v$ be a
 tangent field on $M$ vanishing at the points $c(t_0)$ and
 $c(t_1)$; let $\d_v$ be the prolongation of $v$.

 Along the lifting of $c$ to $TM$, $D$ is $\frac d{dt}$. Then, by
 integrating (\ref{eq:hamilton}) and using that
 $\la\theta,\d_v\ra=\la\theta,v\ra$ vanishes at $t_0$ and $t_1$, we
 find
 \begin{equation}\label{eq:hamilton2}
 \int_{t_0}^{t_1}(\d_vL)\,dt=0\ ,
 \end{equation}
 which is the classical integral form of \emph{Hamilton's
 principle}.

 Newton equation (\ref{gradiente}) is the ``Euler-Lagrange''
 system for the variational principle (\ref{eq:hamilton2}).

 \begin{remarks}\textbf{\em(About constrained systems).}
 \em{
 Let $(M,T_2,\a,\L)$ be a mechanical system with linear constraints
 (\ref{lineal}). Let $\cL\subset TM$ be the linear bundle of
 admissible states, (\ref{admisible}). A vector field $v$ on $M$ is an
 \emph{admissible virtual displacement} when $v_a\in \cL$ for each
$a\in M$; then, the infinitesimal variation $\d_v$ is an
\emph{admissible infinitesimal variation}.

 Such a $\d_v$ is not tangent to $\cL$ in general.
The local uniparametric group generated by $\d_v$ may transform
curves in $\cL$ into curves out of $\cL$, i.e. \emph{kinematically
possible paths} into \emph{kinematically impossible paths}.

 Let $\overline D$ be the second order differential equation
corresponding to $(M,T_2,\a,\L)$.  According (\ref{tiempo}), there
exists  a 1-form $\b$ in $\L$ such that
$$i_{\overline D}\omega_2+dT+\a+\b=0\ .$$

 Then (\ref{eq:zentral})  gives
$$\overline D\la\theta,\d\ra=\d T
        -\la\a,\d\ra-\la\b,\d\ra$$
for each infinitesimal variation $\d$. When $\d=\d_v$ for an
admissible virtual displacement $v$, we have
$\la\b,\d\ra=\la\b,v\ra=0$. When, on top of the above, the system is
also conservative ($\a=dU$), we obtain
\begin{equation}\label{constrained1}
\overline D\la\theta,\d\ra=\d_vL\ ,
\end{equation}
and by the same argument given for (\ref{eq:hamilton2}), we have
\begin{equation}\label{constrained2}
\int_{t_0}^{t_1}(\d_vL)\,dt=0
\end{equation}
when we integrate along a curve $c$ solution of $\overline D$ and
$v$ is any admissible infinitesimal displacement null at $c(t_0)$,
$c(t_1)$.

But now, we cannot in general reach all the kinematically possible
paths close to the given $c$ by means of deformations along
solutions of admissible $\d_v$'s. So, we cannot derive the
extremality of real trajectories among all kinematically possible
paths. See e.g. Whittaker \cite{Whittaker}, pag. 250, for a
discussion.

In the case of holonomic constraints, $\L$ admits local basis of
exact 1-forms $\{dB_1,$ $\dots,dB_r\}$. When $v$ is an admissible
virtual displacement, we have $v(B_k)=0$ and, from Euler commutation
formula, $\d_v\dot B_k=0$. Then, the local equations for $\cL$ in
$TM$: $B_k=\textrm{const}$, $\dot B_k=0$ $(k=1,\dots,r)$ are
preserved by $\d_v$. So each admissible infinitesimal variation is
tangent to $\cL$; and, indeed, tangent to the tangent bundle $TN$
for each one of the solutions $N$ of the distribution $\cL$. In this
way, Hamilton's variational principle specializes, like the Lagrange
equations, in the case of holonomic constraints.

}
\end{remarks}

 \section{Time constraints}

 \begin{definition}\textbf{\em (Time constraint).}
 A \emph{time constraint for} $M$ is a connected closed
hypersurface $\cU$ of $TM$, non intersecting the $0$-section,
projecting regularly onto $M$ and such that the class of time is
locally represented in $\cU$ by exact differential forms.
 \end{definition}

 \begin{remarks}\label{tiempobasico}
 \emph{The last condition means that for each $v_a\in \cU$ there exists
an open neighborhood $\cU_1$ of $v_a$ in $\cU$ and an exact 1-form
$\t_1$ on $\cU_1$ that belongs to the specialization to $\cU_1$ of
the class of time.}

\emph{ Let $f\in\cC^\infty(\cU_1)$ be such that $\t_1=df$. Since the
forms in the class of time are horizontal, so is $df$. Therefore, by
restricting $\cU_1$ to a smaller neighborhood of $v_a$ if necessary,
$f$ is the lift to $\cU_1$ of a function on $M$. So the local
representatives of the class of time in $\cU$ are differentials of
functions on $M$ (lifted to $TM$).}
\end{remarks}

 \begin{lemma} \label{lematiempo1}
 Let  $\cU$ be a locally closed submanifold of $TM$, $\dim\cU=n+r$ ($r\ge
 1$), projecting regularly onto an open set of $M$.
 For each $v_a\in\cU$  there exists a neighborhood of $v_a$ in $TM$ and  $r$
 linearly independent tangent vector fields on this neighborhood,
which are second order differential equations tangent to $\cU$.
 \end{lemma}
 \begin{proof}
 Let $(q^1,\dots, q^n,\dot q^1,\dots,\dot q^n)$ be local coordinates on a
neighborhood of $v_a$ in $TM$. The projection of $\cU$ to $M$ being
regular, and reordering the coordinates if necessary, we can find
local equations
 for $\cU$ of the form:
 $$\dot q^{r+k}=F^{r+k}(q^1,\dots, q^n,\dot q^1,\dots,\dot q^r)
      \qquad (k=1,\dots,n-r)\ .$$

     Then, the fields
 $$D_h=\dot q^i\frac{\partial}{\partial q^i}+
            \frac{\partial}{\partial \dot q^h}+
            \sum_{k=1}^{n-r}\left(
             \dot q^i\frac{\partial F^{r+k}}{\partial q^i}
              +\frac{\partial F^{r+k}}{\partial \dot q^h}
                  \right)
                  \frac{\partial}{\partial \dot q^{r+k}}
        \ ,
          \quad (h=1,\dots, r)$$
 satisfie the requested conditions.
 \end{proof}

 \begin{lemma} \label{lematiempo2}
 Let $\cU$ be as in (\ref{lematiempo1}), and suppose that $\cU$
 does not intersect $0$-section. Given $f\in\cC^\infty(M)$, the differential $df$ represents in $\cU$
the class of time if and only if $\dot f=1$ identically on $\cU$.
 \end{lemma}
 \begin{proof}
 When $\dot f=1$ identically on $\cU$, we have $\frac {df}{\dot f}=df$, so $df$ represents
the class of time.

 Conversely, let $df$ represent the class of time in $\cU$. From (\ref{lematiempo1}),
for $v_a\in\cU$ there exists a second order differential equation
$D$ tangent to $\cU$ in a neighborhood of $v_a$ in $TM$. Because
$df$ belongs to the class of time in $\cU$, we have
 $\langle df, D\rangle=1$ in a neighborhood of $v_a$ in $\cU$. In
 particular, we get
$\langle df, v_a\rangle=1$ and $\dot f(v_a)=1$.
 \end{proof}

 \begin{lemma} \label{lematiempo3}
 Assume $n\ge 2$. Let $\cU$ be a time constraint for $M$. Then, for
$v_a\in \cU$ there exists a neighborhood $U(v_a)$ in $TM$ and
$f\in\cC^\infty(M)$ such that the equation for $\cU\cap U(v_a)$ in
$U(v_a)$ is $\dot f=1$.
 \end{lemma}
 \begin{proof}
 From (\ref{tiempobasico}), there exists $f\in\cC^\infty(M)$ such that
$df$ represents the class of time in a neighborhood of $v_a$ in
$\cU$. From (\ref{lematiempo2}) it follows that $\dot f=1$ in this
neighborhood. We have $d_{v_a}\dot
 f\ne 0$ because $d_af\ne 0$. Thus $\dot f=1$ is the local equation
for an hypersurface of $TM$ in a neighborhood of $v_a$. This
hypersurface has to coincide with $\cU$ in a neighborhood of $v_a$.
 \end{proof}

 \begin{lemma} \label{lematiempo4}
 With the same conditions as in (\ref{lematiempo3}), let $g\in\cC^\infty(M)$
be such that $\dot g=1$ on a neighborhood of $v_a$ in $\cU$. Then
$f$ and $g$ differ by an additive constant on a neighborhood of
$a$ in $M$.
 \end{lemma}
 \begin{proof}
 Let $q^1,\dots,q^n$ be local coordinates for $M$ on a neighborhood of
 $a$, taken such that  $f=q^1$. Then, in the neighborhood
of $v_a$ in $TM$, where the local equation of  $\cU$ is $\dot
 q^1=1$, $\cU$ consists of the tangent vectors to $M$ of the
form
 $$\left(\frac{\partial}{\partial q^1}\right)_b+
     \sum_{j=2}^n\lambda^j \left(\frac{\partial}{\partial
     q^j}\right)_b,
     \quad
     (\lambda^j\in\mathbb{R})\ .$$

 The equation $\dot g=1$ on $\cU\cap U(v_a)$ gives
 $$\frac{\partial g}{\partial q^1}(b)+
     \sum_{j=2}^n\lambda^j \frac{\partial g}{\partial
     q^j}(b)=1$$
 for arbitrary $\lambda^j\in\mathbb{R}$ and $b$ in a neighborhood of $a$.
 This implies $g=q^1+\textrm{const}$ in this neighborhood.
  \end{proof}

 \begin{lemma}\label{lematiempo5}
 Let $\cU$ be  a time constraint for $M$, and $n\ge 2$. For each
$a\in M$, the fiber $\cU_a=\cU\cap T_aM$ is an affine hypersurface
of $T_aM$.
\end{lemma}
\begin{proof}
As $\cU$ is a closed hypersurface of $TM$ that projects regularly
onto $M$, each fiber $\cU_a$ is a closed hypersurface of $T_aM$.
From Lemma (\ref{lematiempo3}) it follows that $\cU_a$ is locally
affine, thus affine.
\end{proof}

  \begin{theorem}\label{teorematiempo}
 There exists a canonical one to one correspondence between closed
1-forms  without zeros on $M$ and time constraints for $M$. The
1-form $\t$ defines the time constraint $\cU$ with equation
$\dot\t=1$ in $TM$, and for each time constraint $\cU$ there is a
unique $\t$ which defines $\cU$ by $\dot\t=1$.
 \end{theorem}
\begin{proof}
 First suppose that $n=\dim M\ge 2$. Let $\cU$ be a time
constraint for $M$. Lemmas (\ref{lematiempo3}) and
(\ref{lematiempo5}) show that for each $a\in M$ there exists a
neighborhood $M_1$ of $a$ in $M$ and a function $f\in\cC^\infty(M)$
such that the equation of $\cU\cap TM_1$ in $TM_1$ is $\dot f=1$.
Lemma (\ref{lematiempo4}) shows that $df$ is uniquely determined by
this condition (reducing $M_1$ if necessary). Patching together the
$df$'s we obtain a closed 1-form $\t$ on $M$ such that $\cU$ is
defined in $TM$ by the equation $\dot\t=1$. And this $\t$ is unique.
Conversely, given $\t$ closed, without zeros on $M$, the equation
$\dot\t=1$ defines a time constraint for $M$.

Now, let $\dim M$ be 1. Then, a closed hypersurface $\cU$ of $TM$
which projects regularly onto $M$ and does not intersect the
$0$-section is the same that a vector field $v$ without zeros on
$M$. The 1-form $\t$ defined by $\la\t, v\ra=1$ defines $\cU$ by
$\dot\t=1$.
\end{proof}

  \begin{remarks}
  \emph{
  According to Theorem (\ref{teorematiempo}), when $M$ is compact and
simply connected, there are no time constraints for $M$, because a
closed 1-form on $M$ is exact, and necessarily has some zeros. In
that case, time constraints can be considered for open submanifolds
of $M$ and non-vanishing exact forms therein.}

\emph{When $M$ is compact, there is no ``time function'', i.e., a
function $t\in\cC^\infty(M)$ such that $dt$ defines a time
constraint, because $dt$ has always zeros. So, if $M$ is compact,
``time has always periods''.}
\end{remarks}

 \section{Modification caused by a time constraint in a free system}

 Let $(M,T_2,\a)$ be a free mechanical system, and $D$ the
 corresponding second order differential equation, so that
 \begin{equation}\label{eq:modification1}
 i_D\omega_2+dT+\a=0
 \end{equation}

 Let $\cU\subset TM$ be the time constraint defined by $\dot\t=1$,
 where $\t$ is a closed 1-form without zeros on $M$.

 Let $\textrm{grad}\,\t$ be the tangent field on $M$ such that
 $i_{\textrm{grad}\,\t}T_2=\t$ (\ref{alfagradiente}), and
 $\textrm{Grad}\,\t$ the vertical field canonically associated to
 $\textrm{grad}\,\t$, so that (Lemma~(\ref{lem:representantevertical}))
 $\displaystyle{i_{\textrm{Grad}\,\t}\o_2=\t}$.

 \begin{theorem}\label{modificationNewton1}
 When $\textrm{grad}\,\t$ is non isotropic for $T_2$ at any point
 of $M$ (in particular when $T_2$ is positive definite), then there
 exists a tangent vector field $\overline D$ on $TM$, satisfying the conditions
 \begin{enumerate}
 \item\label{eq:modif1}
  $i_{\overline D}\o_2+dT+\a\equiv 0\,\,\textrm{mod}\,(\t)$
 \item\label{eq:modif2}
  $\overline D$ is tangent to the hypersurface $\cU$.
 \end{enumerate}
  Such field is a second order differential equation and its
  restriction as tangent field to $\cU$ is uniquely defined by the
  conditions (\ref{eq:modif1}), (\ref{eq:modif2}).
  \end{theorem}

 \begin{proof}
  Let $\overline D$ be defined by
\begin{equation}\label{eq:modification2}
   \overline D=D-
           \frac{D\dot\tau}
              {\langle \textrm{grad}\,\tau,
                      \textrm{grad}\,\tau\rangle}
               \,\textrm{Grad}\,\tau\ ,
\end{equation}
where  
     $\langle\textrm{grad}\,\tau, \textrm{grad}\,\tau\rangle$
 is the scalar product defined by $T_2$.
Inmediately $\overline D$ satisfies (\ref{eq:modif1}), because $D$
satisfies (\ref{eq:modification1}).

Also $\overline D$ satisfies (\ref{eq:modif2}) because we have
$$(\textrm{Grad}\,\tau)(\dot\tau)=
  \langle \textrm{grad}\,\tau,\tau\rangle=
     \langle\textrm{grad}\,\tau, \textrm{grad}\,\tau\rangle\ ,
     $$
thus $\overline D\dot\tau=0$.

Like $D$, $\overline D$ is a second order differential equation,
because $\textrm{Grad}\,\tau$ is vertical.

Any tangent field on $TM$ that satisfies (\ref{eq:modif1}) should
have the form $D+\l\textrm{Grad}\,\tau$ ($\l\in\cC^\infty(TM)$),
and the condition (\ref{eq:modif2}) determines uniquely the value
of $\l$ on $\cU$. That proves the uniqueness of $\overline D$ on
$\cU$.
\end{proof}

Indeed the same  proof shows:

\begin{theorem}\label{modificationNewton2}
 When $\textrm{grad}\,\t$ is non isotropic for $T_2$ at any point
 of $M$ (in particular when $T_2$ is positive definite), there
 exists a tangent vector field $\overline D$ on $TM$, which
 satisfies the conditions
 \begin{enumerate}
 \item\label{eq:modif21}
  $i_{\overline D}\o_2+dT+\a\equiv 0\,\,\textrm{mod}\,(\t)$
 \item\label{eq:modif22}
  $\overline D$ is tangent to each hypersurface $\dot\t=c$ ($c\in\mathbb{R}$)
  of $TM$.
 \end{enumerate}
  Such a field is a second order differential equation and is given
  by formula (\ref{eq:modification2}).
  \end{theorem}

 \begin{remarkN}
 \em{
 According to (\ref{modificationNewton2}), the vector field
 $\overline D$, when restricted to the hypersurface $\dot\t=c$
 ($c\in\mathbb{R}$) gives the evolution equations of the
 mechanical system with time constraint defined by $\frac 1c\,\t$,
 when $c\ne 0$, or holonomic constraint $\dot\t=0$ in the limit
 case $c=0$. In this sense, ordinary holonomic constraints
 result from ``freezing the evolution of the system with respect to $\t$''.}
 \end{remarkN}

 \noindent\textbf{An elementary example.}
 Let $M=\mathbb{R}^3$, $T_2=dx^2+dy^2+dz^2$, $\a=0$ and the time
 constraint $\dot r=1$  ($r=\sqrt{x^2+y^2+z^2}$).

 \noindent\emph{Physical Interpretation:} a moving point is constrained to
 be at a distance $r=t$ from the origin, at each instant $t$,
 with no other external force acting upon it.

 We have
 $$\textrm{grad}\, r=\frac xr\frac\partial{\partial x}
                     +\frac yr\frac\partial{\partial y}
                     +\frac zr\frac\partial{\partial z}\ ,
   \quad \|\textrm{grad}\,r\|=1\ .$$

   Thus, the geodesic field, modified by the time constraint is
 $$\overline D=D-(D\dot r)\textrm{Grad}\,r,\quad
    \text{where}\quad
      D=\dot x\frac\partial{\partial x}
                     +\dot y\frac\partial{\partial y}
                     +\dot z\frac\partial{\partial z}$$
 (the geodesic field for the Euclidean metric on $\mathbb{R}$).

  A direct calculation gives
  $$\overline D=D+
    \left[
     \frac{(x\dot x+y\dot y+ z\dot z)^2}{r^4}-
       \frac{\dot x^2+\dot y^2+\dot z^2}{r^2}\right]
       \left( x\frac\partial{\partial\dot x}
                     + y\frac\partial{\partial\dot y}
                     + z\frac\partial{\partial\dot z}\right)\ ,
   $$
  $\overline D$ is tangent to the manifolds $\dot
  r=\textrm{const}$ in $T\mathbb{R}^3$.

 \noindent i) On $\dot r=0$ (ordinary holonomic constraint $r=\textrm{const}$),
 we have
 $$
 \overline D_0=\dot x\frac\partial{\partial x}
                     +\dot y\frac\partial{\partial y}
                     +\dot z\frac\partial{\partial z}
     -\frac{\dot x^2+\dot y^2+\dot z^2}{r^2}
       \left( x\frac\partial{\partial\dot x}
                     + y\frac\partial{\partial\dot y}
                     + z\frac\partial{\partial\dot z}\right)
 $$
 which is the geodesic field when restricted to each $TS\subset
 T\mathbb{R}^3$, $S=\,\,\,$sphere of radius $r_0$ with the center at the origin.

 In the submanifold $\dot r=0$ of $T\mathbb{R}^3$, to which
 $\overline D_0$ is tangent, the functions $r$ and $v=\sqrt{\dot
 x^2+\dot y^2+\dot z^2}$ are first integrals for $\overline D_0$,
 and the system of differential equations defined by $\overline
 D_0$ is
 $$\frac{d^2x}{dt^2}=-\left(\frac vr\right)^2x\ ,\quad
     \frac{d^2y}{dt^2}=-\left(\frac vr\right)^2y\ ,\quad
      \frac{d^2z}{dt^2}=-\left(\frac vr\right)^2z\ ,$$
which correspond to the classical centripetal force which generate
the motion along geodesics of the sphere.

\noindent ii) On $\dot r=1$ we have
$$
 \overline D_1=\dot x\frac\partial{\partial x}
                     +\dot y\frac\partial{\partial y}
                     +\dot z\frac\partial{\partial z}
     +\frac{1-(\dot x^2+\dot y^2+\dot z^2)}{r^2}
       \left( x\frac\partial{\partial\dot x}
                     + y\frac\partial{\partial\dot y}
                     + z\frac\partial{\partial\dot z}\right)
 $$
that corresponds to a force depending on position and velocity,
which points to or out of the origin according to the value of the
velocity.

 \begin{remarkN}
 \em{
 Later on, we show that, in local coordinates,
 (\ref{eq:modification2}) and $\dot \t=1$ give the
 differential equations for a mechanical system with kinetic
 energy and force depending on time, as they are found in the
 classical literature, e.g. Prange \cite{Prange}, page 556. These
 equations are usually derived from the ``Zentralgleichung'' in
 (\ref{eq:zentral}) when applied to arbitrary $\d$'s \emph{that keep fixed the time},
  i.e. such that $\la\t,\d\ra=0$. But, apparently, there
  appears to be no mathematical justification for this choice of
  $\d$'s, although some authors try to give a physical one; e.g.
  Sommerfeld \cite{Sommerfeld}, page 65, Nordheim \cite{Nordheim},
  pages 48, 49. In that follows, we give a precise mathematical justification,
  starting from the considerartion of time as a constraint.}
 \end{remarkN}

 \begin{lemma}\label{lem:modificada}
 The prolongation $\d_v$ of a vector field $v$ on $M$ to $TM$ is
 tangent to a time constraint $\cU\subset TM$ defined by
 $\dot\t=1$ if and only if $\la\t, v\ra$ is a constant function
 on $M$.
 \end{lemma}
 \begin{proof}
 The problem is local in $M$, so we can take $\t=df$ for some
 $f\in\cC^\infty(M)$.

 Euler commutation formula gives
 \begin{equation}\label{eq:modif3}
 \d_v\dot\t=\d_v\dot f=\dot d(vf)=\dot d\la\t, v\ra\ .
 \end{equation}

 So, $\d_v$ tangent to $\cU$ $\Leftrightarrow$ $\dot d\la\t,
 v\ra=0$ on $\cU$.

 Thus when $\la\t, v\ra$ is a constant function on $M$, $\d_v$
 annihilates $\dot\t$, and is tangent to $\cU$.

 Conversely, when $\d_v$ is tangent to $\cU$, by taking in
 (\ref{eq:modif3}) the value of $\dot d$ at each $w_a\in\cU$, we
 obtain $w_a\la\t, v\ra=0$. But, for each $a\in M$,  the set
 consisting of $w_a\in\cU$ is the affine submanifold of $T_aM$
 defined by $\la w_a,\t\ra=1$, and such submanifold generates
 the whole of $T_aM$. Therefore, $\la\t, v\ra$ is constant on $M$.
 \end{proof}

 \begin{definition}
 \textbf{\em (Admissible infinitesimal displacement).}\label{def:admissible}
 A vector field $v$ on $M$ is an \emph{admissible infinitesimal
 displacement} for the time constraint $\cU$ when $\d_v$ is
 tangent to $\cU$. In that case, $\d_v$ is called an
 \emph{admissible infinitesimal variation}.
 \end{definition}

 From Lemma (\ref{lem:modificada}) it follows

 \begin{theorem}\label{modificadaNewton3}
 Let $(v)$ be the $\cC^\infty(M)$-module generated by the vector
 field $v$ on $M$. The necessary and sufficient condition for
 every field in $(v)$ to be an admissible infinitesimal displacement
 for the time constraint $\cU$ ($\dot\t=1$) is $\la\t, v\ra=0$.
 \end{theorem}

 For variational problems with fixed end points, the admissible
 infinitesimal displacements has to admit multiplication by
 arbitrary functions on $M$. Thus, Theorem
 (\ref{modificadaNewton3}) offers a precise mathematical justification
 for the procedure of ``keeping fixed the time''. \emph{Time must be
 considered as a constraint}.
 \bigskip

 \begin{nada}\textbf{\em Local expressions}
 \end{nada}

 Let us take local coordinates on $M$, $(q^0,q^1,\dots,q^m)$
 ($n=m+1$) with time constraint $\dot q^0=1$; so,
 $q^0$ is \emph{time}. We have
 $\textrm{grad}\,\tau=\textrm{grad}\,q^0$, and
 $\|\textrm{grad}\,\tau\|^2=g^{00}$ (coefficient of the dual
 metric). The differential equations for $(M,T_2,\a)$ with no time
 constraint, with $\alpha=A_i dq^i$, are
  \begin{equation}\label{eq:modificadalocal1}
  g_{ij}\ddot q^j+\Gamma_{hk,i}\dot q^h\dot q^k+A_i=0
  \quad (i=0,1,\dots,m)\ .
  \end{equation}

 When we impose the constraint $\dot q^0=1$,
 the field in (\ref{eq:modificadalocal1}) changes to $\overline D$,
 given by (\ref{eq:modification2}), to which corresponds a force
 form $\overline\a$ that satisfies
 $$\overline\a\equiv\a\,\,\textrm{mod}\,(dq^0)\ .$$

 Therefore changing $D$ by $\overline D$ \emph{does not modify
 Equations (\ref{eq:modificadalocal1}) for indexes} $i\ne 0$ (the
 ``spatial indexes''). By separating spatial indexes
 $\m=1,\dots,m$ from the temporal one $0$, and putting $\dot q^0=1$, we
 obtain the differential equations for the system with $q^0$ as
 time:
  \begin{equation}\label{ecuacionesmodificadas1}
    \begin{cases}
  \dot q^0=1\\
  g_{\mu\nu}\ddot q^\nu+
      \Gamma_{\sigma\nu,\mu}\dot q^\sigma\dot q^\nu+
        2\Gamma_{\nu 0,\mu}\dot q^\nu+
        \Gamma_{00,\mu}
         +A_\mu=0\ ,
  \end{cases}
  \end{equation}
 see e.g. Prange \cite{Prange}, page 556, Eq. (64.a).

  The ``contravariant '' form of the equations results from
  writing (\ref{eq:modification2}) in coordinates. The equations are
 \begin{equation}
    \begin{cases}
  \dot q^0=1\\
  \ddot q^\mu+
      \Gamma_{\sigma\tau}^\mu\dot q^\sigma\dot q^\tau+
        2\Gamma_{\rho 0}^\mu\dot q^\rho+
        \Gamma_{00}^\mu
         +(A^\mu+\frac{D\dot q^0}{g^{00}}\,g^{0\mu})=0\ .
  \end{cases}
  \end{equation}

  Notice that the difference with the equations given by Prange \cite{Prange},
  page 556, Eq. (64.b) is only formal. In Prange's equations the translation from
  subindices to superindices is done by using only the ``spatial''
  matrix $(g_{\m\n})$ instead of the whole matrix $(g_{ij})$ that we use.

  When we use as time $\frac 1c q^0$ instead of $q^0$, the
  equations in the constrained manifold $\dot q^0=c$ are
 \begin{equation}\label{ecuacionesc}
    \begin{cases}
  \dot q^0=c\\
   g_{\mu\nu}\ddot q^\nu+
      \Gamma_{\sigma\nu,\mu}\dot q^\sigma\dot q^\nu+
        2c\,\Gamma_{\nu 0,\mu}\dot q^\nu +
        c^2\,\Gamma_{00,\mu}
         +A_\mu=0\ .
  \end{cases}
  \end{equation}

 When $c=1$, these give (\ref{ecuacionesmodificadas1}). And when
 $c=0$, (\ref{ecuacionesc}) are the equations for the holonomic
 constrained system $(M,T_2,\a,q^0=\textrm{const})$.

\section{Linear constraints depending on time}

 \begin{definition}\textbf{\em (Linear Constraints Depending on
 Time).}\label{def:dependientetiempo}
 Let $(M,T_2,\a)$ be a mechanical system. A system of \emph{linear
 constraints depending on time} is a Pfaff system $\L_M$ on $M$
 together with a closed 1-form $\t$, without zeros on $M$, and
 such that, for each $a\in M$, $\t_a\notin(\L_M)_a$.
 \end{definition}

 A mechanical system with this type of constraints is denoted by
  $(M,T_2,\a,\L_M,\t)$.

 \begin{definition}\textbf{\em (Admissible State).}\label{def:dependientetiempo2}
 An \emph{admissible state} for the constrained mechanical system
 $(M,T_2,\a,\L_M,\t)$ is a point $v_a\in TM$ such that, for each
 $\b_a\in(\L_M)_a$, we have $\la\b_a, v_a\ra=0$, and $\la\t_a, v_a\ra=1$.
 \end{definition}

 The set consisting of all the admissible states is a submanifold
 $\cV\subset TM$, $\cV=\cL\cap\cU$, where $\cL$ is the vector
 distribution annihilated by $\L_M$, and $\cU$ is the time
 constraint defined by $\dot\t=1$. For each $a\in M$, the fiber
 $\cV_a$ is the affine hypersurface of $\cL_a$ defined by equation
 $\la \t_a, v_a\ra=1$.

 \begin{theorem}\textbf{\em (Newton-Lagrange
 Law).}\label{NewtonLagrangedependiente}
 Let $(M,T_2,\a,\L_M,\t)$ be a mechanical system with constraints
 depending on time, and let $T_2$ be positive definite. There exists a
 vector field $\overline D$ on $TM$, which is a second order
 differential equation, satisfying the congruence
 \begin{equation}
 \label{eq:NewtonLagrangedependiente1}
 i_{\overline D}\o_2+dT+\a\equiv 0\,\,\textrm{mod}\,(\L_M,\t)\\
 \end{equation}
 and the \emph{Principle of Virtual Works}:
 \begin{equation}
 \label{eq:NewtonLagrangedependiente2}
 \text{$\overline D$ is tangent to the manifold $\cV$ of admissible
 states.}
 \end{equation}

 The restriction of $\overline D$ to $\cV$ is uniquely defined by
 these conditions.
 \end{theorem}
 \begin{proof}
 Denote by $D$ the second order differential equation
 corresponding to the free system $(M,T_2,\a)$ such that
 \begin{equation}
 \label{eq:NewtonLagrangedependiente3}
 i_D\o_2+dT+\a=0\ .
 \end{equation}

 Let $U$ be a coordinate open set in $M$ such that $\L_M$ admits
 on $U$ a basis $\{\b_1,\dots,\b_r\}$; denote $\t=\b_0$, and
 $v_k=\textrm{grad}\,\b_k$. Let $V_k$ be the vertical
 representative of $v_k$, so that $i_{V_k}\o_2=\b_k$
 ($k=0,1,\dots,r$). Similarly as in Theorem (\ref{teoremaNewtonligaduras}),
 we can find functions $\l^k$ on $TM$ (the ``Lagrange
 multipliers'') such that the vector field $\overline D$ defined on $TU$
 by
 \begin{equation}
 \label{eq:NewtonLagrangedependiente4}
 \overline D=D+\sum_{k=0}^r\l^kV_k\ .
 \end{equation}
 satisfies
 $\overline D\dot\b_0=\overline D\dot\b_1=\dots=\overline
 D\dot\b_r=0$ on $U$.

 $\overline D$ is a second order differential equation and
 satisfies both (\ref{eq:NewtonLagrangedependiente1}) and
 (\ref{eq:NewtonLagrangedependiente2}) on $U$. The field
 $\overline D$ built in this way depends on the choice of the
 local basis $\{\b_1,\dots,\b_r\}$ for $\L_M$, but its restriction
 to the submanifold $\cV\cap TU$ does not. By using an appropriate
 partition of the unity in $M$ (lifted to $TM$) we find a
 $\overline D$ fulfilling (\ref{eq:NewtonLagrangedependiente1}) and
 (\ref{eq:NewtonLagrangedependiente2}) and uniquely defined on
 $\cV$ by these conditions.
 \end{proof}
 \bigskip

 \begin{nada}\textbf{\em Holonomic constraints depending on time.}
 \end{nada}

 Suppose $\L_M$ to be completely integrable. Let $U$ be as in
 the proof of (\ref{NewtonLagrangedependiente}) and
 $\b_k=dB_k$ ($k=0,1,\dots,r)$.

 The equations for $\cV\cap TU$ in $TU$ are
 \begin{equation}\label{eq:holodepende1}
 \dot B_0=1, \dot B_1=\cdots=\dot B_r=0
 \end{equation}

 The field $\overline D$ in (\ref{NewtonLagrangedependiente})
 satisfies
 \begin{equation}\label{eq:holodepende2}
 \overline D\dot B_0=\overline D\dot B_1=\cdots=\overline D\dot B_r=0
 \quad\text{on $\cV\cap TU$.}
 \end{equation}

 Let $N\subset U$ a solution of $\L_M$, given by equations
 \begin{equation}\label{eq:holodepende3}
 B_1=b_1,\dots, B_r=b_r\quad (b_k\in\mathbb{R})\ .
 \end{equation}

 The equations for $\cV\cap TN$ are (\ref{eq:holodepende3}) and
 (\ref{eq:holodepende1}).
 Applying $\overline D$ to equations (\ref{eq:holodepende3}) we
 obtain equalities on $\cV\cap TN$ because of
 (\ref{eq:holodepende1}); and applying $\overline D$ to
 (\ref{eq:holodepende1}) also yields equalities on $\cV\cap TN$
 because of (\ref{eq:holodepende2}). Therefore, $\overline D$ is
 tangent to $\cV\cap TN$.

 For this reason, in the Newton-Lagrange equation
 (\ref{eq:NewtonLagrangedependiente1}) we can specialize  to the
 submanifold $\cV\cap TN$ and obtain
 \begin{equation}\label{eq:holodepende4}
 i_{\overline D_N}\o_{2N}+dT_N+\a_N\equiv 0\,\,\textrm{mod}\,(\t_N)
 \end{equation}
 and
 \begin{equation}\label{eq:holodepende5}
 \overline D\dot\t_N=0\ .
 \end{equation}

 Equations (\ref{eq:holodepende4}) and (\ref{eq:holodepende5})
 show that $\overline D_N$ is the second order differential
 equation for the mechanical system $(N,T_{2N},\a_N)$ with time
 constraint $\dot\t_N=1$. We have proved the following:

 \begin{theorem}\label{NewtonLagrangedependiente2}
 Let $(M,T_2,\a)$ be a mechanical system with $T_2$ positive
 definite. Let $\L_M$ be a completely integrable Pfaff system on
 $M$. Let $\t$ be a closed 1-form without zeros on $M$, $\dot\t=1$ the time
 constraint defined by $\t$. Let $\overline D$ be the second order
 differential equation corresponding to the system
 $(M,T_2,\a,\L_M,\t)$. Then, for each submanifold $N\subset M$
 solution of $\L_M$, $\overline D$ is tangent to the time
 constraint $\dot\t_N=1$ in $TN$, and the restriction of
 $\overline D$ to this submanifold is the second order
 differential equation corresponding to
 $(N,T_{2N},\a_N,\t_N)$ by the Newton-Lagrange law
 (\ref{modificationNewton1}).
 \end{theorem}

 \begin{remarkN}
 \em{
 Theorem (\ref{NewtonLagrangedependiente2}) means
 that time constraints \emph{specialize} properly in the presence of holonomic
 constraints: we can introduce the time constraint after
 specializing $T_2$ and $\a$ to the submanifolds solutions of
 $\L_M$.}
 \end{remarkN}
 \begin{remarkN}
 \em{
 It is clear that the precedent methods can be applied in some cases in which
 $T_2$ is not positive definite, under appropriate hypothesis
 (e.g. when $\textrm{grad}\,\t$ is a field of the distribution
 $\cL$ and the restrictions of $T_2$ to $\cL$ and to the
 orthogonal complement of $\cL$ are positive or negative
 definite).}
 \end{remarkN}

\section{Reference frames and inertial forces}\label{c:inertial}

\begin{nada}\textbf{\em General principles.}
\end{nada}
In  the context of Lagrangian Mechanics it is natural to consider as
\emph{reference frame} of a manifold $\cM$ a manifold isomorphism
$\varphi\colon\cR\to\cM$ that transports mechanical structures from
$\cM$ to $\cR$, where we ``read them''. When $\cR$ and $\cM$ are
provided with pseudo-Riemannian  metrics, the geodesic field of
$\cM$, once transported to $\cR$, is a second order differential
equation that, by substraction of the geodesic field of $\cR$, gives
a vertical field on $T\cR$: the \emph{inertial force caused by}
$\varphi$. The most relevant dynamical elements appear when $\cR$
and $\cM$ are endowed with time constraints which correspond to each
other by $\varphi$.

Uniparametric automorphisms groups of a given manifold $M$ are a
particular case when we consider one such group as an automorphism
of the manifold $\cM=\mathbb{R}\times M$. Classical examples are
uniparametric groups of isometries of $\mathbb{R}^3$, which generate
centrifugal and Coriolis forces. The same method allow us to deal
with more general groups, like the one of dilatations that are
apparently ignored in the literature.

\begin{example}\textbf{\em (Inertial forces caused by automorphism
groups)}\label{ex:inertialforces}
\end{example}

 Let $\varphi\colon\cR\to\cM$ be an isomorphism of manifolds. Let
 $T_2$ be a metric on $\cM$, and $\theta$ the correspondig Liouville
 form (see (\ref{remark:covariant})). It is easy to prove that
 $\varphi^*\theta$ is the Liouville form for $\varphi^*T_2$ on
 $\cR$; thus, if $\o_2=d\theta$, $\varphi^*\o_2$ is the 2-form
 associated to $\varphi^*T_2$ on $\cR$.

 In particular, let $T_2$ be non degenerate (i.e., a
 pseudo-Riemannian metric) and $\varphi$ be an isomorphism. Then,
 $\varphi_*$ applies the second order differential equation $D$
 into $\overline D$, related by:
 $$i_D(\varphi^*\o_2)+d(\varphi^*T)+\varphi^*(\overline\a)=0$$
 when
 $$i_{\overline D}\o_2+dT+\overline\a=0\ .$$

 When $\overline\a=0$, we observe that the geodesic field for
 $\varphi^*T_2$ on $\cR$ is transformed into the geodesic field for
 $T_2$ on $\cM$. When $\cR$ is also equipped with a metric, the
 vertical tangent field on $T\cR$, difference between the geodesic
 field for $\varphi^*T_2$ and the one of the metric given on $\cR$,
 is the \emph{inertial force produced by} $\varphi$.

 Let us consider the case $\cR=\cM=\mathbb{R}\times M$, where $M$
 is an $n$-dimensional manifold.
 And let $\varphi\colon\mathbb{R}\times M\to\mathbb{R}\times M$
 an uniparametric automorphism group:
 $$\varphi(t,a)=(t,\varphi_t(a))\ ,$$
where, for each $t\in\mathbb{R}$, $\varphi_t\colon M\to M$ and
$\varphi_0=\textrm{Id}$, $\varphi_t\circ\varphi_s=\varphi_{t+s}$, as
usual. Denote by $u$ the infinitesimal generator of $\varphi$.

 Given a metric $T_2$ on $M$, we endow $\mathbb{R}\times M$ with
the metric $\widetilde{T}_2=dt^2\oplus T_2$. We give a formula for
$\varphi^*\widetilde{T}_2$.

 We have $\varphi_*\left(\frac\partial{\partial t}\right)=
    \frac\partial{\partial t}+u$ (we use the product structure on
$\mathbb{R}\times M$ to extend $u$ from $M$ to $\mathbb{R}\times
M$, and $\frac\partial{\partial t}$ from $\mathbb{R}$ to
$\mathbb{R}\times M$). Thus
$$\left(\varphi^*\widetilde{T}_2\right)_{(t,a)}\left(
               \frac\partial{\partial t},
                  \frac\partial{\partial t}\right)=
      1+\| u\|^2_{\varphi_t(a)}\ .$$

  For a field $v$ on $M$,
\begin{align*}
\left(\varphi^*\widetilde{T}_2\right)_{(t,a)}\left(
               \frac\partial{\partial t},
                  v\right)
          &=\widetilde{T}_2\left(
               \frac\partial{\partial t}+u,
                  \varphi_*v\right)_{(t,\varphi_t(a))}
            =T_2(u_{\varphi_t(a)},\varphi_{t*}v_a)
            =\varphi_t^*T_2(u_a,v_a)\\
          &=\la i_{u_a}(\varphi_t^*T_2),v_a\ra
           =\left(dt\otimes i_{u_a}(\varphi_t^*T_2)\right)
                  \left(\frac\partial{\partial t},v_a\right)\\
          &=\left(2\,dt\, i_{u_a}(\varphi_t^*T_2)\right)
                \left(\frac\partial{\partial t},v_a\right)\ .
\end{align*}

 For two fields $v$, $w$ on $M$:
$$\left(\varphi^*\widetilde{T}_2\right)_{(t,a)}(v,w)=
    \widetilde{T}_2(\varphi_{t*}v_a,\varphi_{t*}w_a)=
     \left(\varphi_t^*\widetilde{T}_2\right)_a(v,w)\ .$$

 We obtain:
\begin{equation}\label{eq:inertial1}
\left(\varphi^*\widetilde{T}_2\right)_{(t,a)}=
         \left(1+\|u\|^2_{\varphi_t(a)}\right)\,dt^2
          +2\,dt\,i_{u_a}\left(\varphi^*_tT_2\right)
              +\left(\varphi^*_tT_2\right)_a\ .
\end{equation}

To illustrate with simple numerical examples, let us take
$M=\mathbb{R}^2$, $T_2=dx^2+dy^2$; then
$\widetilde{T}_2=dt^2+dx^2+dy^2$. And the most classical groups:
\medskip

\noindent\textbf{Translations:}
 $\varphi\colon (t,x,y)\mapsto (t,x+t,y)$.

In this case, we have $u=\partial/\partial x$ and
\begin{equation}\label{eq:inertial2}
\varphi^*\widetilde T_2=2\,dt^2+2\,dt\,dx+dx^2+dy^2\ .
\end{equation}
\medskip

\noindent\textbf{Rotations:}
   $\varphi\colon(t,x,y)\mapsto (t,x\cos t-y\sin t,
x\sin t+y\cos t)$.
$$u=-y\frac\partial{\partial x}+x\frac\partial{\partial y}\ ,
    \quad \|u\|^2=x^2+y^2\ ,
    \quad \varphi^*_tT_2=T_2\ ,$$
thus, from (\ref{eq:inertial1}), we obtain
\begin{equation}\label{inertial3}
 \varphi^*\widetilde T_2=
     (1+x^2+y^2)\, dt^2+2\,dt(-ydx+xdy)+dx^2+dy^2\ ,
\end{equation}
that can also be calculated directly.
\medskip

 \noindent\textbf{Dilatations:}
  $\varphi\colon(t,x,y)\mapsto (t,e^tx,e^ty)$,
$$u=x\frac\partial{\partial x}+y\frac\partial{\partial y}\ ,
    \quad \|u\|^2=x^2+y^2\ ,
    \quad \varphi^*_tT_2=e^{2t}T_2\ ,$$
$$\|u\|^2_{\varphi_t(x,y)}=e^{2t}(x^2+y^2)\ ,
    \quad i_u(\varphi^*_tT_2)=e^{2t}(xdx+ydy)\ ,$$
thus
\begin{equation}\label{eq:inertial4}
\varphi^*\widetilde T_2=[1+e^{2t}(x^2+y^2)]\, dt^2+
     2\,e^{2t}\,dt(xdx+ydy)+e^{2t}(dx^2+dy^2)\ .
\end{equation}

We can use these formulas in order to calculate the inertial forces
associated to each of these groups. In each case, we consider in
$\mathbb{R}\times M$ the time constraint $\dot t=1$ ($\mathbb{R}$ is
the clock for the system). The geodesic field for the given metric
$\widetilde T_2$ is $\cD_G=\dot t\partial/\partial t
           +\dot x\partial/\partial x
            +\dot y\partial/\partial y,$
which is tangent to the given constraint.
\medskip

 \noindent\textbf{Translations.} The transformed metric
  $\varphi^*\widetilde T_2$ has constant coefficients. Then
all Christoffel symbols are zero, and the geodesic field for
$\varphi^*\widetilde T_2$ is the same as the one for $\widetilde
T_2$. The inertial force is 0.
 \medskip

 \noindent\textbf{Rotations.} A routine tedious computation for
the field $\cD$ such that
$$i_\cD(\varphi^*\widetilde\o_2)+d(\varphi^*\widetilde T)=0$$
($\cD$ transforms by $\varphi_*$ to the geodesic field on
 $T(\mathbb{R}\times \mathbb{R}^2)$) gives
$$\cD=\dot t\frac\partial{\partial t}
                     +\dot x\frac\partial{\partial x}
                     +\dot y\frac\partial{\partial y}
     +(x\,\dot t^2+2\,\dot t\dot y)\frac\partial{\partial\dot x}
        +(y\,\dot t^2-2\,\dot t\dot x)\frac\partial{\partial\dot y}
                       $$
which, in the locus of the time constraint $\dot t=1$ is
\begin{equation}\label{eq:inertial5}
\cD|_{\dot t=1}=\frac\partial{\partial t}
                     +\dot x\frac\partial{\partial x}
                     +\dot y\frac\partial{\partial y}
     +(x+2\,\dot y)\frac\partial{\partial\dot x}
        +(y-2\,\dot x)\frac\partial{\partial\dot y}\ .
 \end{equation}

 The difference with the geodesic field is the inertial force
corresponding to the group of rotations. The differential
equations for this force are
\begin{equation}\label{eq:inertial6}
\frac{d^2x}{dt^2}=x+2\,\dot y\ ,\quad
 \frac{d^2y}{dt^2}=y-2\,\dot x\ ,
\end{equation}
where we observe the sum of a centrifugal and a Coriolis force.
\medskip

\noindent\textbf{Dilatations.} A similar computation gives for the
field $\cD$ which transforms by $\varphi_*$ into the geodesic field,
when restricted to time constraint $\dot t=1$:
\begin{equation}\label{eq:inertial7}
\cD|_{\dot t=1}=\cD_G
     -(x+2\,\dot x)\frac\partial{\partial\dot x}
        -(y+2\,\dot y)\frac\partial{\partial\dot y}\ .
 \end{equation}

 To which corresponds the differential equations for the inertial
force caused by $\varphi$:
\begin{equation}\label{eq:inertial8}
\frac{d^2x}{dt^2}=-(x+2\,\dot x)\ ,\quad
 \frac{d^2y}{dt^2}=-(y+2\,\dot y)\ .
\end{equation}

We do not give a physical interpretation of this formal result.

\begin{definition}\textbf{\em (Inertial frames).}\label{def:inertial frames}
 Let $\cR$, $\cM$ be pseudo-Riemannian manifolds. A reference
frame $\varphi\colon\cR\to\cM$ is called \emph{inertial} when the
force caused by $\varphi$ is zero. This applies in particular to
uniparametric automorphism groups.
\end{definition}

Equivalent characterizations of inertial frames are: $\varphi$ is
inertial if and only if $\varphi_*$ transforms the geodesic field of
$T\cR$ into the geodesic field of $T\cM$. And also, if and only if
it transforms one to another the Levi-Civita connections on $\cR$
and $\cM$.
\bigskip

\begin{example}\textbf{\em (Inertial groups on $\mathbb{R}^n$).}\label{ex:inertialgroups}
\end{example}

The metric we consider in $\mathbb{R}^n$ is the usual Euclidean one.
We prove that the only inertial uniparametric groups on
$\mathbb{R}^n$ are the translation ones.

Retaining the notations of (\ref{ex:inertialforces}), for $\varphi$
to be inertial, $\widetilde{T}_2$ and $\varphi^*\widetilde{T}_2$
must define the same Levi-Civita connection on $\mathbb{R}\times
\mathbb{R}^n$. Thus, when written in cartesian coordinates, all the
Christoffel symbols for $\varphi^*\widetilde{T}_2$ must to be zero;
and then, the $g_{ij}$ for $\varphi^*\widetilde{T}_2$ are constants.
Looking at the last term in (\ref{eq:inertial1}), this gives
$\varphi_t^*{T}_2=\varphi_0^*{T}_2=T_2$. Then, since the coefficient
in the second term in (\ref{eq:inertial1}) are constant, we obtain
that $u$ has constant coefficients. We conclude that $\varphi$ is
the translation group generated in $\mathbb{R}^n$ by $u$,
{\em{q.e.d}}.
\bigskip

\begin{nada}\textbf{\em Preservation of the equations of motion}.\label{preservation}
\end{nada}

In the physical literature, inertial frames are frequently referred
as those which ``preserve the form of the equations of motion'',
without defining a precise meaning to ``preserve''. We treat this
question in this section.

 Let $(\cR,T_2)$ and $(\cM,\overline T_2)$ be pseudo-Riemannian
 manifolds, and  $\varphi\colon\cR\to\cM$ a reference frame ($=$
 manifold isomorphism). Let $\overline D$ be a second order
 differential equation on $T\cM$, and $\overline\a$ its canonically
 associated 1-form:
 \begin{equation}\label{eq:inertial9}
 i_{\overline D}\overline\o_2+d\overline T+\overline\a=0
 \end{equation}

 There are two different natural ways to define a second order
 differential equation on $T\cR$ corresponding by $\varphi$ to
 $\overline D$:

 \noindent {\bf a)} The second order differential equation $D$ on $T\cR$
 canonically associated to $\varphi^*(\overline\a)$:

 \begin{equation}\label{eq:inertial10}
 i_D\o_2 + dT +\varphi^*(\overline\a)=0\ .
 \end{equation}
 \medskip

 \noindent {\bf b)} The field $D_1$ on $T\cR$ such that $\varphi_*D_1=\overline
 D$:

 \begin{equation}\label{eq:inertial11}
 i_{D_1}(\varphi^*\overline\o_2) + d(\varphi^*\overline T) +\varphi^*(\overline\a)=0\ .
 \end{equation}

 Classically, $D_1$ is the field obtained from $\overline D$ by
 \emph{change of coordinates}.

 The difference $D_1-D$ is a vertical field on $T\cR$, i.e., a
 force; when we think of $D_1$ as $\overline D$ ``read in the coordinates of
 $\cR$'', $D_1-D$ is \emph{the force caused by} $\varphi$ \emph{in
 the mechanical system} $(\cM,\overline T_2,\overline\a)$. The
 frame may be inertial in the sense of Definition~(\ref{def:inertial
 frames}), but causes a non zero force when $\overline\a\ne 0$, as
 shown in the following
 \begin{proposition}\label{prop:inertialframes}
 Let $(\cR,T_2)$ and $(\cM,\overline T_2)$ be pseudo-Riemannian
 manifolds, and $\varphi\colon\cR\to\cM$ a reference frame. The force
 caused by $\varphi$ in $(\cM,\overline T_2,\overline\a)$ is zero for all choices
 of $\overline\a$, if and only if $\varphi$ is an
 isometry.
 \end{proposition}
 \begin{proof}
 The ``if'' part is obvious. For the converse, let us take
 $\overline\a=0$; then, for the geodesic field $D_G$ on $T\cR$, we
 have
 \begin{equation}\label{eq:inertial12}
 i_{D_G}\o_2 + dT=0,\quad
 i_{D_G}(\varphi^*\overline\o_2) + d(\varphi^*\overline T)=0\ .
 \end{equation}

 Then, let $\overline\a$ be any horizontal 1-form on $T\cM$, and
 $\a=\varphi^*(\overline\a)$; let $D$ be given by
 (\ref{eq:inertial10}), and define the vertical field $V$ by
 $D=D_G+V$, so that, from
 (\ref{eq:inertial10},\ref{eq:inertial12}),
 $$i_V\o_2+\a=0\ .$$

 By taking the field $D_1$ in (\ref{eq:inertial11}), the hypothesis
 $D_1=D$ gives, from (\ref{eq:inertial11},\ref{eq:inertial12}),
 $$i_V(\varphi^*\overline\o_2)+\a=0\ .$$

  Thus $i_V\o_2=i_V(\varphi^*\overline\o_2)$; then, the geometric
  representative $v$ of $V$ satisfies
  $$i_vT_2=i_v(\varphi^*\overline T_2)\ .$$

 Since $v$ is arbitrary, this implies $T_2=\varphi^*\overline T_2$.
 \end{proof}

 Thus it is not  true that the uniparametric groups of
 translations considered in (\ref{ex:inertialforces}) do not
 produce forces; they do, but these forces may be compensated by
 time constraints, as we will show.

 \begin{definition}\textbf{\em (Preservation of the equations of
 motion).}\label{def:preservation}
 Let $(\cR,T_2)$ and $(\cM,\overline T_2)$ be pseudo-Riemannian
 manifolds with respective time constraints $\dot\t=1$ and $\dot{\overline\t}=1$.
 Let $\varphi\colon\cR\to\cM$ be a reference frame.
 We say that $\varphi$ \emph{preserves the equations of motion}
 when $\varphi^*(\overline\t)=\t$ and, for each second order
 differential equation $\overline D$ on $T\cM$ such that
 $\overline D\dot\t=0$, and for the corresponding $\overline\a$ by
 (\ref{eq:inertial9}), and for the field $D_1$ on $T\cR$ such that
 $\varphi_*D_1=\overline D$, we have
 \begin{equation}\label{eq:inertial13}
 i_{D_1}\o_2+dT+\varphi^*(\overline\a)\equiv
 0\,\,\textrm{mod}\,(\t)\ .
 \end{equation}
 \end{definition}

  $D_1$ is tangent to the time constraint $\dot\t=1$, and the
  congruence (\ref{eq:inertial13}) means that $D_1$ is the
  modification of $D$ in (\ref{eq:inertial10}) caused by the
  time constraint, in the sense of Theorem
  (\ref{modificationNewton1}).

  \begin{theorem}\label{th:inertial}
  An uniparametric automorphism group
  $\varphi\colon \mathbb{R}\times M\to\mathbb{R}\times M$
  preserves the equations of motion if and only if it is inertial
  and also each $\varphi_t\colon M\to M$ is an isometry.
  \end{theorem}
  \begin{proof}
   The problem reduces to prove that for an inertial
   $\varphi$, the condition of preserving the equations of motion
   is equivalent to be an isometry group of $M$.

   For any uniparametric automorphism group $\varphi$ the geodesic
   field $D_G$ satisfies $D_G\dot t=0$.

   Let $\overline D=D_G+\overline V$ be a second order
   differential equation on $T(\mathbb{R}\times M)$ such that
   $\overline D\dot t=0$, i.e. $\overline V\dot t=0$. We keep
   notations from (\ref{ex:inertialforces}) and define
   $\widetilde\a$ by
   $$i_{\overline D}\widetilde\o_2+d\widetilde T+\widetilde\a=0,\quad
   \text{or}\quad i_{\overline V}\widetilde\o_2+\widetilde\a=0\ .$$

  Let $D_1$ be the field defined by $\varphi_*D_1=\overline D$,
  and put $D_1=D_G+V_1$. As $\varphi$ is inertial, we have
  $\varphi_*V_1=\overline V$, thus
  \begin{equation}\label{eq:inertial14}
  i_{V_1}(\varphi^*\widetilde\o_2)+\varphi^*(\widetilde\a)=0\ .
  \end{equation}

   Let $D=D_G+V$ be defined by
  \begin{equation}\label{eq:inertial15}
  i_D\widetilde\o_2+d\widetilde T+\varphi^*(\widetilde\a)=0\ ,
  \end{equation}
  thus
  \begin{equation}\label{eq:inertial16}
  i_V\widetilde\o_2+\varphi^*(\widetilde\a)=0
  \end{equation}
  and from (\ref{eq:inertial14}),
  \begin{equation}\label{eq:inertial17}
  i_V\widetilde\o_2=i_{V_1}(\varphi^*\widetilde\o_2)\ .
  \end{equation}

  The condition for $\varphi$ to preserve the equations of motion
  is
  \begin{equation}\label{eq:inertial18}
  i_{D_1}(\widetilde\o_2)+d\widetilde T+\varphi^*(\widetilde\a)\equiv
  0\,\,\textrm{mod}\, (dt)
  \end{equation}

  Substracting (\ref{eq:inertial15}) from (\ref{eq:inertial18}),
  this condition is
  $$i_{V_1}(\widetilde\o_2)
      -i_V(\widetilde\o_2)\,\equiv\,
  0\,\,\textrm{mod}\, (dt)$$
  or, from (\ref{eq:inertial17})
  $$i_{V_1}(\widetilde\o_2)\,\equiv\,
  i_{V_1}(\varphi^*\widetilde\o_2)\,\,\textrm{mod}\, (dt)\ .$$

  By taking geometric representatives, this condition is
  \begin{equation}\label{eq:inertial19}
  i_{v_1}(\widetilde T_2)\, \equiv\,
  i_{v_1}(\varphi^*\widetilde T_2)\,\,\textrm{mod}\, (dt)\ .
  \end{equation}

  Thus the necessary and sufficient condition for $\varphi$ to
  preserve the equations of motion is that (\ref{eq:inertial19})
  holds for each tangent field $v_1$ on $T(\mathbb{R}\times M)$
  taking values in $T(\mathbb{R}\times M)$ and satisfying $v_1t=0$
  (this is $V_1\dot t=0$).

  Looking at (\ref{eq:inertial1}) we see that this condition for
  $\varphi$ is equivalent to have $\varphi^*_tT_2=T_2$ for each
  $t\in\mathbb{R}$.
   \end{proof}
  \begin{corollary}\label{cor:inertial}
  The uniparametric automorphism groups of $\mathbb{R}^n$ which
  preserve the equations of motion are only the inertial groups,
  i.e., the translation uniparametric groups.
  \end{corollary}

\section*{Acknowledgements.}

The author is grateful to Prof. Ricardo Alonso and Prof. Ricardo
P\'erez Marco for useful and inspiring mathematical discussions. The
author, TeX illiterate, also thanks R. Alonso for all of his time
and generous work in the typesetting of this article; and to both
Ricardo's for their help with the English.


\providecommand{\bysame}{\leavevmode\hbox
to3em{\hrulefill}\thinspace}

\end{document}